\documentclass[prb,twocolumn,showpacs,preprintnumbers,amsmath,amssymb]{revtex4}

\usepackage{graphicx}
\usepackage{bm}

\begin{document}
\title
{
A New Method of the High Temperature Series Expansion
}
\author
{
Noboru {\sc Fukushima}\footnote{new address will be:
Institut f\"{u}r Theoretische Physik,
Technische Universit\"{a}t Braunschweig,
38106 Braunschweig, Germany\\
e-mail: n.fukushima@tu-bs.de
 }
}

\affiliation
{
Max-Planck-Institut f\"ur Physik komplexer Systeme,
N\"othnitzer Stra{\ss}e 38, D-01187 Dresden, Germany
}
%

\begin{abstract}
We formulate a new method of performing high-temperature
series expansions for the spin-half Heisenberg model or,
more generally,
for SU($n$) Heisenberg model with arbitrary $n$.
The new method is a novel extension of the well-established
finite cluster method. Our method emphasizes hidden combinatorial
aspects of the high-temperature series expansion,
and solves the long-standing problem of how to
efficiently calculate correlation functions of operators
acting at widely separated sites.
Series coefficients are expressed in terms of cumulants,
which are shown to have the property that all deviations
from the lowest-order nonzero cumulant can be
expressed in terms of a particular kind of moment expansion.
These ``quasi-moments'' can be written in terms
of corresponding ``quasi-cumulants'', which enable us to
calculate higher-order terms in the high-temperature series
expansion.
We also present a new technique for obtaining
the low-order contributions to specific heat from
finite clusters.

\bigskip

{\bf keywords}: high-temperature expansion, Heisenberg model,
correlation function, specific heat, new method
%
%
%
%
\end{abstract}

\maketitle

\section{Introduction}

The high-temperature expansion can be used in any dimension
and has provided significant information on variety of models.\cite{Domb3}
It is based on the Taylor expansion of
the Boltzmann factor $e^{-\beta {\cal H}}$ in $\beta$,
around the high-temperature limit.
Although the concept of the high-temperature expansion is quite simple,
its high-order calculation requires sophisticated strategies.
A standard method of the high-temperature expansion is
called the finite cluster method\cite{Domb3,oitmaa,Gelfand} (FCM).
It has been used for a long time,\cite{Domb3}
and progress of the high-temperature series has been
due mainly to improvement of computers.
In the FCM, the series expansion is reduced to
calculation in connected finite-size clusters.
However,
the FCM is inefficient
for calculating a correlation function between rather distant sites:
(i) The clusters have to include those two sites
and calculation in such large clusters consumes a lot of time.
(ii) The FCM does not use a valuable piece of information.
That is, the high-temperature series of a long-range correlation
starts at a certain finite order, namely,
low-order coefficients are equal to zero -----
physically, it means that a long-range correlation
develops at low temperature.

In this paper, we formulate a new method
oriented to the long-range correlation,
namely, it overcome (i) and (ii) above.
The new method is valid for the spin-1/2 Heisenberg model,
or, more generally,
for SU($n$) Heisenberg model with arbitrary $n$.
High-temperature series coefficients are written in terms of cumulants.
We start from the lowest-order nonzero cumulant
in the correlation-function series.
Next, we consider a deviation from it.
Then, the deviation can be
regarded as a sort of moment, which we call a quasi-moment.
The corresponding quasi-cumulant enables us to calculate
a number of terms higher than the lowest-order nonzero cumulant.
In Ref.~\onlinecite{FK1d},
the Fourier transform of the correlation function is
calculated using the new method and the FCM complementarily.
That is, we have used
the new method for the long-range correlations
and
the FCM for the short-range correlations.

In addition, also for the specific heat,
we have adopted a similar strategy in Ref.~\onlinecite{FK1d}.
That is, contributions from the required largest cluster are calculated
by a new technique, and
the FCM is used only for the smaller clusters.
We utilize cumulants relevant to the ordering of quantum variables.
By choosing nonzero cumulants from them,
some of the series coefficients are calculated simply.
We explain also this new technique in this paper.

The structure of this paper is as follows.
In Sec.~\ref{sec:model}, we present the model.
We show a representation of SU($n$) Heisenberg model
in terms of exchange operators.
The relation of a correlation function and
an expectation value of the exchange operator is mentioned.
In Sec.~\ref{sec:htse}, we review the high-temperature expansion.
Although we use this model for illustration,
what is written here can generally be applied to other models.
To clarify the purpose of our new method,
we review the FCM in detail
using a one-dimensional system as an explicit example.
In Sec.~\ref{sec:newcor}, we describe
the new method for the correlation function.
Here, we derive some properties of the cumulants.
Owing to those properties, we can define quasi-moments/cumulants.
How to calculate series coefficients using those quasi-cumulants
is explained.
In Sec.~\ref{sec:newsh},
we show the new technique for the specific heat.
Finally, we give a summary in Sec.~\ref{sec:con}.

\section{model}
\label{sec:model}

Let us consider the spin-1/2 Heisenberg model,
\begin{equation}
 {\cal H}^{(p)}_{\rm SU(2)}:= 2 \sum_{(i,j)}
J_{i,j} \; \bm{s}_{i} \cdot \bm{s}_{j},
\label{hamilsu2}
\end{equation}
where $\bm{s}_i$ is the spin operator at site $i$, and
$J_{i,j}$ is the coupling constant.
Here, $(i,j)$ represents
that the summation is performed only once for each $(i,j)$ pair.
The spin operators are rewritten as
\begin{equation}
2 \; \bm{s}_{i} \cdot \bm{s}_{j} = P_{i,j}-\frac12,
\end{equation}
where $P_{i,j}$ is an exchange operator defined by
\begin{eqnarray}
P_{i,j}|\cdots \stackrel{i}{\alpha} \cdots \stackrel{j}{\beta} \cdots \rangle
= |\cdots \stackrel{i}{\beta}\cdots\stackrel{j}{\alpha}\cdots \rangle,
\label{eq:exchangeop}
\end{eqnarray}
for arbitrary $\alpha=\uparrow,\downarrow$, and $\beta=\uparrow,\downarrow$.
Then, we can rewrite Eq.~(\ref{hamilsu2}) as
\begin{equation}
 {\cal H}^{(p)}_{\rm SU(2)}= \sum_{(i,j)}J_{i,j} \left(P_{i,j}-\frac12
\right).
\end{equation}
This Hamiltonian has SU(2) symmetry.
Since our method is valid more generally for an SU($n$) symmetric case
with arbitrary positive integer $n$,
we define the SU($n$) Heisenberg model below.

Let each site take one of the $n$ colors,
and denote them as $|\alpha\rangle$ with $\alpha = 1, 2, \cdots, n$.
An exchange operator is defined
using $X^{\alpha\beta}:= |\alpha\rangle\langle\beta|$
by
\begin{equation}
 P_{i,j}:= \sum_{\alpha=1}^n  \sum_{\beta=1}^n
X_i^{\alpha\beta} X_j^{\beta\alpha},
\end{equation}
for $i\neq j$.
Colors at sites $i$ and $j$ are exchanged by $P_{i,j}$,
namely, Eq.~(\ref{eq:exchangeop}) is satisfied
for arbitrary colors $\alpha$ and $\beta$.
Furthermore, for a later convenience,
let us define
\begin{equation}
p_{i,j}:=P_{i,j}-\frac1{n}.
\end{equation}
In the case of SU(2),
\begin{equation}
p_{i,j}=2 \; \bm{s}_{i} \cdot \bm{s}_{j}
\label{eq:su2spexp}
\end{equation}
is satisfied.
Although we mostly use expressions for general $n$ in this paper,
we expect that Eq.~(\ref{eq:su2spexp})
helps instant understanding of
explanations.
In some cases, we explicitly use the SU(2) notation
for a simple explanation.

The Hamiltonian of the SU($n$) Heisenberg model is given by
\begin{equation}
 {\cal H}^{(p)}:= \sum_{(i,j)}J_{i,j} \left(P_{i,j}-\frac1{n}
\right)
= \sum_{(i,j)}J_{i,j} p_{i,j}.
\end{equation}
The constant term just yields an energy shift.
Hence, for calculating thermal averages, we can use
\begin{equation}
 {\cal H}:=  \sum_{(i,j)}J_{i,j} P_{i,j}.
\label{hamil}
\end{equation}
We define
the partition function as
\begin{equation}
 Z := {\rm Tr}\; e^{-\beta {\cal H}},
\label{eq:parfunc}
\end{equation}
and the free energy as
\begin{equation}
F :=- \beta^{-1} \ln Z.
\end{equation}
The average in this system is denoted by
\begin{equation}
\langle \hat{O} \rangle :=
{\rm Tr} \left(\hat{O} \; e^{-\beta {\cal H}}\right) /Z,
\end{equation}
where $\hat{O}$ is an arbitrary operator.

In the case of spin-1/2 system, there are
three independent interacting components,
$s^x, s^y, s^z$.
In general, when there are $n$ states per site,
the maximum number of independent interacting components is $n^2-1$.
The model, isotropic with respect to these components,
is the SU($n$) Heisenberg model
as explicitly shown in Appendix \ref{app:derive}.

What we present in this paper is the high temperature
expansion of $\langle P_{i,j} \rangle$.
A correlation function
$\langle X_i^{\alpha\beta} X_j^{\beta\alpha} \rangle$
is calculated using a relation,
\begin{eqnarray}
\langle X_i^{\alpha\beta} X_j^{\beta\alpha} \rangle
&=&
\frac{1}{n^2-1} \left( \langle  P_{i,j} \rangle -\frac{1}{n} \right),
\label{eq:x2p}
\end{eqnarray}
for arbitrary $\alpha \neq \beta$, $i\neq j$.
In addition, when $i= j$,
\begin{equation}
\langle X_i^{\alpha\beta} X_i^{\beta\alpha} \rangle
=\frac{1}{n},
\end{equation}
for arbitrary $\alpha \neq \beta$.
These relations are shown also in Appendix \ref{app:derive}.
In the case of SU(2),
$\langle X_i^{\alpha\beta} X_j^{\beta\alpha} \rangle=
 2 \langle s_i^{z} s_j^{z} \rangle$.
The uniform susceptibility is equal to inverse temperature times
the Fourier transform of the correlation function
with wave-number zero.
Furthermore,
the specific heat is calculated from
the internal energy
$\langle {\cal H} \rangle= \sum_{(i,j)} J_{i,j}\langle P_{i,j} \rangle$.

\section{high-temperature expansion}
\label{sec:htse}

In this section,
we review high-temperature expansion.
Although the Heisenberg model is used for explanation,
this review is more general.
That is,
one can easily replace the Heisenberg model with a different model.

\subsection{Moments and cumulants for classical variables}
\label{sec:mcclassic}

Fundamental properties of moments and cumulants are frequently
used in this paper.
In order to conveniently refer to those properties,
let us review moments and cumulants of classical variables
$x_i \; (i=1,2,\cdots)$.
We put most of the fundamental details of the review
in Appendix \ref{appn:fdclvbl},
and let us here just show one important formula used frequently in this paper.
Here,
a moment is denoted by $\langle x_j x_k \cdots x_l \rangle_x$,
and a cumulant is denoted by $[ x_j x_k \cdots x_l ]_x$.

Moments can be expanded using cumulants, and vice versa.
However, in those relations, the number of terms in the expansion
drastically increases as the order of moments or cumulants
becomes higher. To avoid it,
another relation,
\begin{eqnarray}
[ x_i  \cdots x_\ell ]_{x}
& = & \langle x_i  \cdots x_\ell \rangle_{x} \nonumber \\
& & - \sum_{{\cal P}(\xi,2)}
[ x_i \cdots x_j ]_{x}
\langle x_k \cdots x_m \rangle_{x},
\label{eq:mixexpand}
\end{eqnarray}
can be used.\cite{Domb3}
Here, $\xi$ is the number of $x$-variables in the bracket in the l.h.s.
Then, the summation denoted by ${\cal P}(\xi,2)$
is taken over every partition of $\xi$ elements into two groups
on the condition that
one of the variables, for example $x_i$,
must always be included in $[ \cdots ]_{x}$.
In addition, each bracket includes at least one $x$-variable.
In this paper,
we call this formula ``the mixed expansion pivoting on $x_i$''.
For example, the mixed expansion of $[ x_1 x_2 x_3 ]_x$
pivoting on $x_1$ is written as
%
\begin{eqnarray*}
 [ x_1 x_2 x_3 ]_x &=&
\langle x_1 x_2 x_3 \rangle_{x} -
[x_1 x_2 ]_{x} \langle x_3 \rangle_{x} -
[x_1 x_3 ]_{x} \langle x_2 \rangle_{x}  \\ &&-
[x_1]_{x} \langle x_2 x_3\rangle_{x} .
\end{eqnarray*}

\subsection{Moments and cumulants for quantum variables}
\label{sec:cum4qnum}

As a reference system, we take a non-interacting system such that $J_{i,j}=0$
for every $(i,j)$ pair.
Its partition function is written as $Z_0:={\rm Tr}\,1 $, and
the average in the non-interacting system is denoted by
\begin{equation}
\langle \hat{O} \rangle_0 := \frac{{\rm Tr}(\hat{O})} {{\rm Tr}\, 1},
\end{equation}
where $\hat{O}$ is an arbitrary operator.
This average $\langle \cdots \rangle_0$ plays a role of
$\langle \cdots \rangle_x$ in the previous subsection.

There are a number of ways to define
moments and cumulants of quantum variables
because of their noncommutativity.\cite{Kubo}
One of the ways is defining a moment
by a symmetrized product,
%
\begin{eqnarray}
&& \langle P_{i_1,j_1} P_{i_2,j_2}\cdots
 P_{i_k,j_k}\rangle_{\rm s}
\nonumber\\
&& := \frac{1}{k !} \sum_\sigma
\langle P_{i_{\sigma(1)},j_{\sigma(1)}} P_{i_{\sigma(2)},j_{\sigma(2)}}\cdots
 P_{i_{\sigma(k)},j_{\sigma(k)}}\rangle_0, \quad
\label{eq:symmetrization}
\end{eqnarray}
where $\sigma$ represents a permutation of the indeces.
Note that the ordering of exchange operators $P_{i,j}$ is unimportant
in $\langle\cdots\rangle_{\rm s}$.
Hereafter, we use
$\langle\cdots\rangle_0$ rather than $\langle\cdots\rangle_{\rm s}$
if they are obviously equivalent to each other,
{\it e.g.},
$\langle P_{i,j} \rangle_{\rm s}= \langle P_{i,j} \rangle_0 $;
further examples are commented in Appendix \ref{appn:sym}.

The partition function, Eq.~(\ref{eq:parfunc}), is rewritten as
\begin{equation}
 Z={\rm Tr}\,1 \; \frac{{\rm Tr}\; e^{-\beta {\cal H}}}{{\rm Tr}\,1}
  = Z_0 \; \langle e^{-\beta {\cal H}} \rangle_0.
\label{eq:avebol}
\end{equation}
This averaged Boltzmann factor works as
the generating function of the symmetrized moments mentioned above,
namely, with $\lambda_{i,j}=-\beta J_{i,j}$,
\begin{equation}
\langle P_{i,j} \cdots
 P_{k,l} \rangle_{\rm s}=
\left.
\frac{\partial}{\partial \lambda_{i,j}}
\cdots
\frac{\partial}{\partial \lambda_{k,l}}
 \langle e^{- \beta {\cal H}} \rangle_{0}
\right|_{\lambda=0}.
\end{equation}
The corresponding cumulants are denoted by $[ \cdots ]_{\rm s}$,
and given by
\begin{equation}
[ P_{i,j} \cdots P_{k,l} ]_{\rm s}=
\left.
\frac{\partial}{\partial \lambda_{i,j}}
\cdots
\frac{\partial}{\partial \lambda_{k,l}}
\ln \langle e^{- \beta {\cal H}} \rangle_{0}
\right|_{\lambda=0}.
\label{eq:symcumdef}
\end{equation}
These moments and cumulants have
the properties explained in the previous subsection and
Appendix \ref{appn:fdclvbl}.

The free energy is rewritten as
\begin{equation}
F = - \frac{1}{\beta} \ln Z_0
 -\frac{1}{\beta} \sum_{m=1}^\infty \frac{(-\beta)^{m}}{m!}
[{\cal H}^m]_{\rm s},
\label{eq:hmcs0}
\end{equation}
%
\begin{eqnarray}
[{\cal H}^m]_{\rm s} &=&
\sum_{(i_1,j_1)} \cdots \sum_{(i_{m},j_{m})}
J_{i_1,j_1}\cdots J_{i_{m},j_{m}}
\nonumber \\ &&
\qquad\qquad\qquad \times
   \left[
P_{i_1,j_1} \cdots P_{i_{m},j_{m}}
   \right]_{\rm s}.
\label{eq:hmcs}
\end{eqnarray}
%
Many terms in the summation of Eq.~(\ref{eq:hmcs}) are
equivalent
because the ordering of operators in $[\cdots]_{\rm s}$ is unimportant.
That is,  in the case of the cumulant
$[P_{1,2}^{k_{12}} P_{2,3}^{k_{23}} \cdots ]_{\rm s}$,
then, $m!/(k_{12}!k_{23}!\cdots)$ terms are equivalent.

From Eqs.~(\ref{eq:hmcs0}) and (\ref{eq:hmcs}), we can derive
\begin{eqnarray}
& & \langle P_{i,j}\rangle
  =
\frac{\partial F}{\partial J_{i,j}}  \nonumber \\
& & \quad =
 \sum_{m=1}^\infty \frac{(-\beta)^{m-1}}{(m-1)!}
\sum_{(i_1,j_1)} \cdots \sum_{(i_{m-1},j_{m-1})}
\nonumber \\ & & \quad\qquad
J_{i_1,j_1}\cdots J_{i_{m-1},j_{m-1}}
   \left[ P_{i,j}
P_{i_1,j_1} \cdots P_{i_{m-1},j_{m-1}}
   \right]_{\rm s} \nonumber \\
& &\quad=
  \sum_{m=0}^\infty \frac{(-\beta)^m}{m!}
   \left [P_{i,j}{{\cal H}}^m \right]_{\rm s}.
\label{eq:Wheikin}
\end{eqnarray}
Even if $J_{i,j}=0$ in the original Hamiltonian
({\it e.g.}, for non-nearest neighbors in the case of
nearest-neighbor interaction),
one can start from the general formulation written above.
After deriving all the formulae,
one can put $J_{i,j}=J$ or 0
according to the original Hamiltonian.

Let us consider cumulants for ${p}_{i,j}$ instead of $P_{i,j}$.
See the definition of cumulants, Eq.~(\ref{eq:symcumdef}).
If one replaces $P_{i,j}$ by ${p}_{i,j}$,
the generating function of cumulants changes only by
$-1/n \sum_{(i,j)} \lambda_{i,j}$, and thus
only the first-order cumulants are affected by this replacement.
In other words,
\begin{equation}
[ P_{i,j} ]_{\rm s} = \langle P_{i,j}\rangle_0  =1/n
\label{eq:1stcumP}
\end{equation}
differs from
\begin{equation}
[{p}_{i,j}]_{\rm s}= \langle {p}_{i,j}\rangle_0  =0.
\label{eq:1stcumXX}
\end{equation}
On the other hand, the higher order cumulants are equal.
Namely,
\begin{equation}
[{p}_{i_1,j_1}{p}_{i_2,j_2}\cdots
{p}_{i_\ell,j_\ell}]_{\rm s}=
[P_{i_1,j_1} P_{i_2,j_2}\cdots
 P_{i_\ell,j_\ell}]_{\rm s}
\label{eq:cumequal}
\end{equation}
when $[\cdots]_{\rm s}$ has two or more ${p}_{i,j}$ operators.
For practical calculation,
we mainly use $P_{i,j}$ for convenience.
In some cases, however,
the property $\langle{p}_{i,j}\rangle_0=0$ helps simplification.
Hereafter, we use $P_{i,j}$ and ${p}_{i,j}$
interchangeably with keeping in mind
Eqs.~(\ref{eq:1stcumP})-(\ref{eq:cumequal}).

\subsection{Properties of the cumulants}

Let us introduce a diagrammatic representation of
the moments and the cumulants.
Although we mainly use those diagrams for later calculation,
let us define them here in advance because
they are convenient also for simple explanation in this review.

\begin{figure}[h]
\includegraphics[width= 7cm]{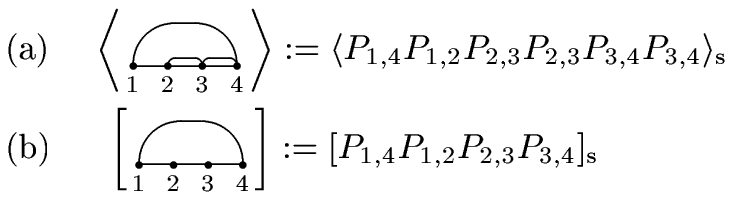}
\caption{ Examples of bond-diagrams.
\label{fig:bonddia1a}}
\end{figure}

Figure \ref{fig:bonddia1a} shows
examples of the diagrams for $\langle \prod P_{i,j} \rangle_{\rm s}$
and $[ \prod P_{i,j}]_{\rm s}$.
We call them bond-diagrams.
Each dot represents a site.  A segment connecting sites $i$ and $j$
represents $P_{i,j}$, and we call it a bond.
Note that
a product of two equivalent operators appearing in
$\langle\cdots\rangle_{\rm s}$
[{\it e.g.}, $P_{3,4}P_{3,4}$ in the r.h.s$.$ of Fig.~\ref{fig:bonddia1a}(a)]
cannot be reduced to the identity because these are
{\it not} always next to each other in the symmetrization.
Figure \ref{fig:bonddia1a}(a) is a {\it moment} bond-diagram,
and (b) is a {\it cumulant} bond-diagram.

\begin{figure}[h]
\includegraphics[width= 6cm]{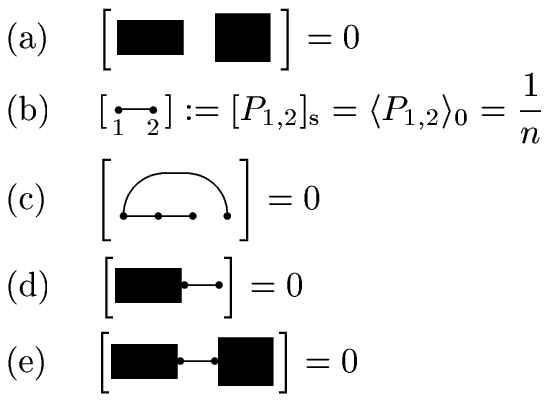}
\caption{
Properties of the cumulants.
A solid rectangle represents arbitrary bonds.
\label{fig:bonddia1b}}
\end{figure}

A cumulant is equal to zero if the variables in  $[\cdots]_{\rm s}$
can be partitioned into two groups
which are
independent of each other on averaging $\langle \cdots \rangle_0$.
For example, $[P_{1,2} P_{3,4}]_{\rm s}=0$
because the trace of sites 1,2 is taken
independently of that of sites 3,4
in the non-interacting system.
In other words,
a cumulant is equal to zero if the bond-diagram is not {\it connected}
as shown in Fig.~\ref{fig:bonddia1b}(a).
Here, a solid rectangle in Fig.~\ref{fig:bonddia1b}
represents arbitrary bonds.
Therefore, as for cumulants, we consider only connected diagrams hereafter.

Figure \ref{fig:bonddia1b}(b) is a first-order cumulant.
In a higher-order cumulant bond-diagram,
more than one bonds must meet at each dot.
If not, it is equal to zero as shown in
Fig.~\ref{fig:bonddia1b}(c) and (d).
Furthermore,
a cumulant bond-diagram is equal to zero
if two parts in it are connected by only one bond
as shown in Fig.~\ref{fig:bonddia1b}(e).
We prove these properties
in Appendix \ref{app:addprf}.

\subsection{An example of the finite cluster method}
\label{sec:exFCM}

The finite cluster method (FCM) is
a standard method for the high-temperature expansion.\cite{Domb3,oitmaa}
It is also used for perturbative expansions starting from
exactly solvable models.\cite{Gelfand}
In order to explain it in detail,
we give an explicit example
using the one-dimensional nearest-neighbor interacting
system with a periodic boundary condition, namely,
$J_{i,j}=J \delta_{j,i+1}$, $(1\le i \le N)$, where the site $N+1$
is equivalent to site 1.
In this case, the procedure of the FCM
is reduced to a very simple form.

In the FCM,
series coefficients in the thermodynamic limit are exactly obtained
by summing contributions from finite-size clusters.
The clusters are subsets of the infinite lattice, and
we need only connected clusters
because cumulant bond-diagrams must be ``connected''.
Namely, the one dimensional system requires only open chains.
The cluster of size $\ell$ is defined as
\begin{equation}
{\cal H}_{\ell}^{(p)}
:= J \sum_{i=1}^{\ell-1}\left(P_{i,i+1}-\frac1{n}\right) =
J \sum_{i=1}^{\ell-1} p_{i,i+1},
\label{eq:FCMHamiltonian}
\end{equation}
where we have chosen to use $p_{i,j}$ rather than $P_{i,j}$
in order to make use of $[{p}_{i,j}]_{\rm s}=0$.
The expectation value in this cluster is given by
\begin{eqnarray}
\langle {p}_{i,j}\rangle_\ell
& := &  \frac{{\rm Tr}( {p}_{i,j} e^{-\beta{\cal H}_{\ell}^{(p)}} )}
{{\rm Tr} (e^{-\beta{\cal H}_{\ell}^{(p)}})} \nonumber \\
& = &
  \sum_{m=0}^\infty \frac{(-\beta)^m}{m!}
 \left[{p}_{i,j}\big({\cal H}_{\ell}^{(p)}\big)^m\right]_{\rm s}.
\end{eqnarray}
We assume $i<j$ hereafter.
Note that
$\langle {p}_{i,j}\rangle_\ell = 0$ when $i<1$ or $j>\ell$
because of a property shown in Fig.~\ref{fig:bonddia1b}(c) and (d).
Since a finite cluster is used instead of the original system,
$\langle {p}_{i,j}\rangle_\ell$ contains only a subset of terms
appearing in $\langle {p}_{i,j}\rangle$.
The important point is that $\langle {p}_{i,j} \rangle_\ell$
includes contributions from clusters smaller than $\ell$, and thus
simple summation of $\langle {p}_{i,j} \rangle_\ell$ over $\ell$
yields multiple counting of those terms.
To avoid this, the FCM requires
subtraction of contributions from the smaller clusters.
We define a {\it net} contribution from the $\ell$-site cluster,
%
\begin{eqnarray}
\langle {p}_{i,i+x}\rangle_\ell^\prime
& := &
\langle {p}_{i,i+x}\rangle_\ell \nonumber \\ & & -
\sum_{\ell_1=0}^{i-1}\mathop{{\sum}'}_{\ell_2=0}^{\ell-j}
\langle {p}_{(i-\ell_1),(i-\ell_1)+x}\rangle_{\ell-\ell_1-\ell_2}^\prime,
\label{eq:excsum}
\end{eqnarray}
where the summation ${\sum}'$ excludes $\ell_1=\ell_2=0$.
Here, $\ell_1$ ($\ell_2$) represents the number of removed sites
from the side of 1 ($\ell$).
When $i=1$ and $i+x=\ell$, this expression is reduced to
$\langle {p}_{1,\ell}\rangle_{\ell}^\prime =
\langle {p}_{1,\ell}\rangle_{\ell}$.
Let $\langle {p}_{i,i+x}\rangle'_{\le\ell}$ denote
the total net contribution to $\langle {p}_{i,i+x}\rangle$
from clusters smaller than or equal to $\ell$,
\begin{eqnarray}
\langle {p}_{i,i+x}\rangle'_{\le\ell}:=
\sum_{l=x+1}^\ell
\sum_{j=1}^{l-x} \langle {p}_{j,j+x} \rangle_l^\prime.
\label{eq:totcont}
\end{eqnarray}
Here, $\langle {p}_{i,i+x}\rangle'_{\le\ell}$ does not depend on
$i$ because of the translational symmetry.
Note that
$\langle {p}_{i,i+x}\rangle=\langle {p}_{i,i+x}\rangle'_{\le\infty}$.
When $\ell$ is finite, the series
for $\langle {p}_{i,i+x}\rangle'_{\le\ell}$
is correct up to a finite order.
To think about the ``finite order'',
let us take a cumulant bond-diagram appearing in the $m$-th order.
The diagram has a bond $P_{i,i+x}$
and $m$ bonds from the Hamiltonian.
In order to give a nonzero contribution,
there must be at least one bond per nearest-neighbor pair between $i$ and $j$.
Furthermore,
nearest-neighbor pairs not between $i$ and $i+x$
----- less than $i$ or greater than $i+x$ -----
must have at least two bonds per pair.
Therefore, there is a contribution from $(\ell+1)$-site cluster
when $m\geq x + 2 (\ell-x)$.
That is,
$\langle {p}_{i,i+x}\rangle'_{\le\ell}$
is correct up to $O[(\beta J)^{ 2 \ell -x -1}]$.

The above is the usual method of the FCM.
However, in one dimension, these equations can be reduced to
simpler forms as shown in Appendix \ref{app:freeene}.

\section{A new method for the correlation function}
\label{sec:newcor}

Let us think about calculating
the Fourier transform of the correlation function,
\begin{equation}
S^z(q):=\frac{1}{n^2-1}\sum_{x} \langle p_{j+x,j} \rangle e^{iqx}.
\end{equation}
Series of $S^z(q)$ up to $O[(\beta J)^M]$ requires
series of each $\langle p_{j+x,j} \rangle$ up to $O[(\beta J)^M]$.
Here, the important point is that
the series coefficients of $\langle p_{j+x,j} \rangle$
up to $O[(\beta J)^{|x|-1}]$
are equal to zero as explained in Fig.~\ref{fig:bonddia1b}(c),(d) and
in Sec.~\ref{sec:exFCM}.
Hence,
we need $\langle p_{j+x,j} \rangle$ only for $1\le x \le M$.
In addition, that property provides us a weak point of
the FCM: When $M$ is fixed,
large $x$ requires a larger cluster than small $x$ does.
Thus, the most time-consuming part is calculation in the largest cluster,
$x=M$, namely, calculation of
Tr$^{(1+M)} \{p_{1,1+M} ({\cal H}^{(p)}_{1+M})^M\}$.
However, what we need for $x=M$ is
only the $M$-th order coefficient
because we already know that the lower orders
are equal to zero.
Therefore, if only the FCM is used,
the most time-consuming calculation yields
very little information.

Our goal in this paper to formulate a new method is as follows.
If the contributions from large clusters
can be calculated by another method,
the FCM can be used only for smaller clusters.\cite{Singhcomment}
Accordingly, one can calculate up to high orders.
In Ref.~\onlinecite{FK1d}, we have calculated $S^z(q)$
up to $O[(\beta J)^{19}]$.
However, we have used the FCM only for $\ell\le 13$.
The corrections to the results have been calculated by the method
explained in this section.

\subsection{Traces using combinatorics}
\label{sec:trace}

We calculate traces of products of exchange operators by
decomposing every permutation
into a product of independent cyclic permutations\cite{hand,Chen}
as explained in the following,
(an explicit example is given in Appendix \ref{sec:xtrace}).
Let us consider a trace of
$P:=P_{i_1, j_1}P_{i_2, j_2}\cdots P_{i_m, j_m}$,
with ${\rm Tr}^{(\ell)}$ denoting the trace in the $\ell$-site system,
\begin{eqnarray}
&& {\rm Tr}^{(\ell)}P \nonumber \\
&:=&
\sum_{\alpha_1=1}^n  \cdots \sum_{\alpha_\ell=1}^n
\langle\stackrel{1}{\alpha_1}\stackrel{2}{\alpha_2}\cdots\stackrel{\ell}{\alpha_\ell} |
P
|\stackrel{1}{\alpha_1}\stackrel{2}{\alpha_2}\cdots\stackrel{\ell}{\alpha_\ell} \rangle
\nonumber
\\
& =&
\sum_{\alpha_1=1}^n  \cdots \sum_{\alpha_\ell=1}^n
\langle\stackrel{1}{\alpha_1}\stackrel{2}{\alpha_2}\cdots\stackrel{\ell}{\alpha_\ell} |
\stackrel{1}{\alpha_{P1}}\stackrel{2}{\alpha_{P2}}\cdots
\stackrel{\ell}{\alpha_{P\ell}} \rangle
\nonumber
\\
& =&
\sum_{\alpha_1=1}^n  \cdots \sum_{\alpha_\ell=1}^n
\delta_{\alpha_1,\alpha_{P1}}
\delta_{\alpha_2,\alpha_{P2}}
\cdots
\delta_{\alpha_\ell,\alpha_{P\ell}}
,
\label{trace}
\end{eqnarray}
where $\alpha_{Pi}$ refers to the color at position $i$ after
the permutation $P$.
The summation of $\alpha_i$ makes a contribution
only when $\alpha_i=\alpha_{Pi}$ for every $i$.
Consider using this relation successively starting from $i$.
That is, $\alpha_i$ is equal to $\alpha_{Pi}$, and then
$\alpha_{Pi}$ is equal to $\alpha_{P^2i}$, \dots,
one can repeat this procedure until coming back to
$\alpha_i$ at a certain power of $P$,
namely,
 $\alpha_i=\alpha_{Pi}=\alpha_{P^2i}=\cdots=\alpha_i$.
In other words,
all the variables whose subscript belong to one cyclic permutation in $P$
have to be equal.
Since any permutation can be decomposed into a product of independent
cyclic permutations,
the number of independent variables of the summation
is the number $Y(P)$ of cyclic permutations of $P$.
Therefore the trace is given by,
${\rm Tr}^{(\ell)} P =n^{Y(P)}$, and accordingly,
\begin{equation}
 \langle P \rangle_\ell =
\frac{{\rm Tr}^{(\ell)} P }{{\rm Tr}^{(\ell)} 1}=n^{Y(P)-\ell}.
\label{trace2}
\end{equation}

In fact, this method allows us to
calculate Tr$^{(\ell)} (P_{i,j} P) $ at the same time
as Tr$^{(\ell)} P $ for
an arbitrary $(i,j)$ pair.\cite{Chen}
Let us remember
the permutation $P$ is decomposed into cycles
in calculating ${\rm Tr}^{(\ell)} P$.
When the cycle which $i$ belongs to is different from
what $j$ belongs to,
the next operation $P_{i,j}$ unites
the two cycles into one.
Then, the number of cycles decreases by one,
namely, ${\rm Tr}^{(\ell)} (P_{i,j}P)=n^{Y(P)-1}$.
On the contrary,
if $i$ and $j$ belong to one cycle of $P$,
then $P_{i,j}$ breaks the cycle into two.
Then, the number of cycles increases by one,
namely, ${\rm Tr}^{(\ell)} (P_{i,j}P)=n^{Y(P)+1}$.

\subsection{Explicit calculation of cumulants}
\label{sec:amida-diagrams}

\begin{figure}[h]
\includegraphics[width= 7cm]{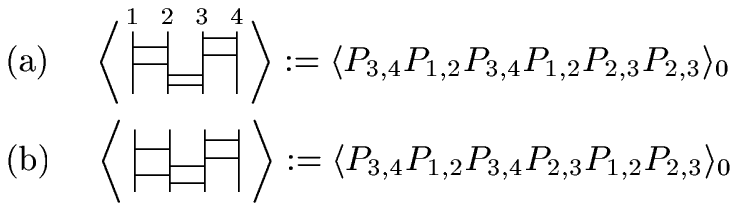}
\caption{Examples of amida-diagrams.
\label{fig:defdia}}
\end{figure}

To represent $\langle\prod P_{i,j} \rangle_0$,
we introduce an unsymmetrized version of a bond-diagram.
Figure \ref{fig:defdia} shows its examples.
Since this diagram is equivalent to an ``amida lottery'',
used in Japan to peacefully decide
how to distribute a fixed number of objects among
an equal number of people,
here we call it an amida-diagram.
Each vertical line represents a site, and
a horizontal line between site $i$ and $j$ represents
$P_{i,j}$.
The order of horizontal lines from bottom to top
should coincide with the order of
$P_{i,j}$ in $\langle \cdots \rangle_0$ from right
to left.
We bracket an amida-diagram by $\langle\cdots\rangle$
in order to distinguish it from
a cumulant version of the amida-diagram
introduced later in Sec.~\ref{sec:newsh}.

\begin{figure}[h]
\includegraphics[width= 4cm]{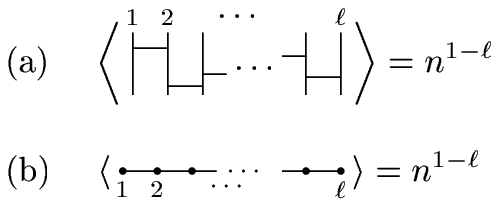}
\caption{ Actual calculated values of diagrams.
The dotted line represents a sequence of bonds
such that there is only one bond between sites.
\label{fig:property}}
\end{figure}

Since expectation values are not changed
by transpositions of neighboring commutative operators,
the ordering of $P_{1,2}$ and $P_{3,4}$ is irrelevant to the results.
On the other hand,
two exchange operators sharing one of the site-indeces, such as
$P_{1,2}$ and $P_{2,3}$, are not commutative,
and in general the ordering of those bonds affects the result.
For example, the value of Fig.~\ref{fig:defdia}(a) is different from
that of (b).
However,
in some cases, even the order of such ``site-sharing'' bonds
does not affect the result.
Let us think about the value of an amida-diagram
in Fig.~\ref{fig:property}(a)
by counting cycles.
The ``initial'' state is the identity permutation,
which has $\ell$ cycles.
Then, every time $P_{i,j}$ is applied,
it unites two cycles into one.
After all the $P_{i,j}$ operators are applied,
there is only one cycle, which includes all the $\ell$ sites.
Hence the expectation value is $n^{1-\ell}$,
and it does not depend on the ordering of those $P_{i,j}$ operators.
Therefore, after the symmetrization,
the moment bond-diagram Fig.~\ref{fig:property}(b) also gives $n^{1-\ell}$.

Let us think about adding $P_{1,\ell}$ to
Fig.~\ref{fig:property}(a) on the top.
Since sites 1 and $\ell$ belong to the same cycle,
the operation of $P_{1,\ell}$ cuts the cycle.
Namely, the permutation has two cycles and
the diagram gives $n^{2-\ell}$.
Again, after the symmetrization,
the corresponding moment bond-diagram also gives $n^{2-\ell}$.

\begin{figure}[h]
\includegraphics[width= 8cm]{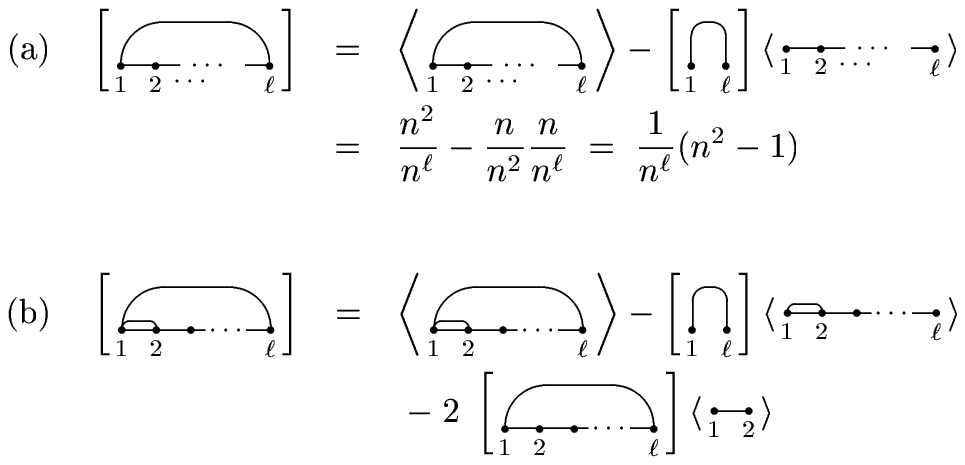}
\caption{The mixed expansion pivoting on $P_{1,\ell}$
for (a) the lowest-order and
(b) the second-lowest-order nonzero cumulants
in the $\langle {p}_{1,\ell}\rangle$ series.
The dotted line represents a sequence of bonds
such that there is only one bond between sites.
\label{fig:bonddia2}}
\end{figure}

In Fig.~\ref{fig:bonddia2}(a),
we calculate a cumulant using the values derived above.
This is the lowest-order nonzero cumulant
in the $\langle {p}_{1,\ell}\rangle$ series.
We use the mixed expansion, Eq.~(\ref{eq:mixexpand}),
pivoting on $P_{1,\ell}$.
The expression of the expansion is very simple
because $[P_{1,\ell}]_{\rm s}$ is the one and only
nonzero cumulant which includes $P_{1,\ell}$.

\begin{figure}[h]
\includegraphics[width= 8cm]{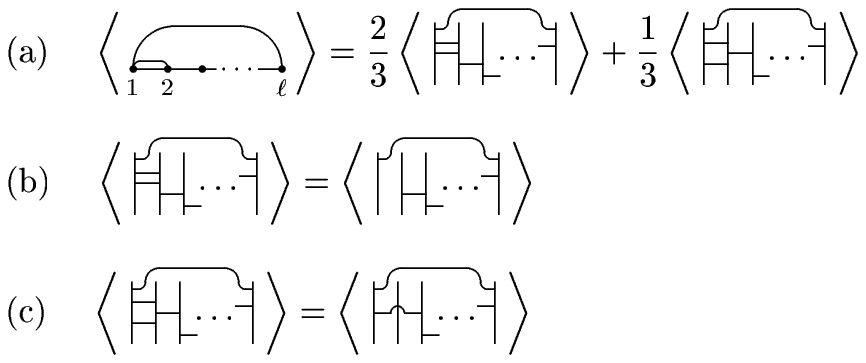}
\caption{Calculation of a symmetrized moment using amida-diagrams.
\label{fig:extrabond}}
\end{figure}

Next, let us consider adding one more bond, $P_{1,2}$.
The mixed expansion pivoting on $P_{1,\ell}$
is shown in Fig.~\ref{fig:bonddia2}(b).
The coefficient 2 of the last term is
due to the two ways of choosing
which one of the two $P_{1,2}$ is put in $[\cdots]_{\rm s}$.
%
%
The first term of the r.h.s$.$ is calculated
as shown in Fig.~\ref{fig:extrabond}.
The symmetrization is equivalent to averaging
over all the possible configurations of bonds.
In fact, however,
the value of each amida-diagram is determined only by
relative configurations of two $P_{1,2}$ and one $P_{2,3}$.
There are only two patterns, namely, Fig.~\ref{fig:extrabond}(b) and (c).
Their r.h.s$.$ are obtained by simplification
$P_{1,2}P_{1,2}=1$ and $P_{1,2}P_{2,3}P_{1,2}=P_{1,3}$.
Here, an ``$\Omega$-shape'' bond in Fig.~\ref{fig:extrabond}(c)
represents that the bond is not connected to the site at ``$\Omega$''.
We already know how to calculate the r.h.s$.$ of
Fig.~\ref{fig:extrabond}(b) and (c).
We also know that configuration of $P_{i,i+1}$ ($i\ge 3$)
does not affect the value of the amida-diagram.
That is, the value of the amida-diagram is determined only by
the position of $P_{2,3}$.

In order to obtain the l.h.s$.$ of Fig.~\ref{fig:extrabond}(a),
we have to count how often each of the configurations
Fig.~\ref{fig:extrabond}(b) and (c)
appears in the symmetrization.
For example,
the prefactor 1/3 in the last term of Fig.~\ref{fig:extrabond}(a)
is obtained by
\begin{equation}
\frac
{\{\mbox{the number of configurations of Fig.~\ref{fig:extrabond}(c)}\}}
{\{\mbox{the total number of possible configurations}\}}.
\label{eq:onethird0}
\end{equation}
It can be considered like this:
The vertical line of site 2 is partitioned into three regions
by the two $P_{1,2}$.
The probability of finding $P_{2,3}$ in the intermediate region is 1/3.
More explicitly, it can be given also by integrals,
\begin{equation}
\frac{
\int_0^{1}  {\rm d}\tau_{12}
\int_{\tau_{12}}^1  {\rm d}\tau_{12}^{\prime}
\int_{\tau_{12}}^{\tau_{12}^{\prime}}  {\rm d}\tau_{23} }
{
\int_0^{1}  {\rm d}\tau_{12}
\int_{\tau_{12}}^1  {\rm d}\tau_{12}^{\prime}
\int_{0}^{1}  {\rm d}\tau_{23} }
=\frac13,
\label{eq:onethird}
\end{equation}
where $\tau_{12}$, $\tau_{12}^{\prime}$, $\tau_{23}$
represent positions of
lower $P_{1,2}$, upper $P_{1,2}$, $P_{2,3}$,
respectively.
Here, $\tau=0$ is the bottom, $\tau=1$ is the top.
On the other hand,
the ``probability'' of the other configuration
Fig.~\ref{fig:extrabond}(b) is 2/3.
Consequently,
the symmetrized value is calculated as shown in Fig.~\ref{fig:extrabond}(a).
This formula is true only when $\ell \geq 3$.
If $\ell=2$, we cannot take an average over position of $P_{2,3}$,
and Fig.~\ref{fig:extrabond}(c) does not appear.

\begin{figure}[h]
\includegraphics[width= 7.5cm]{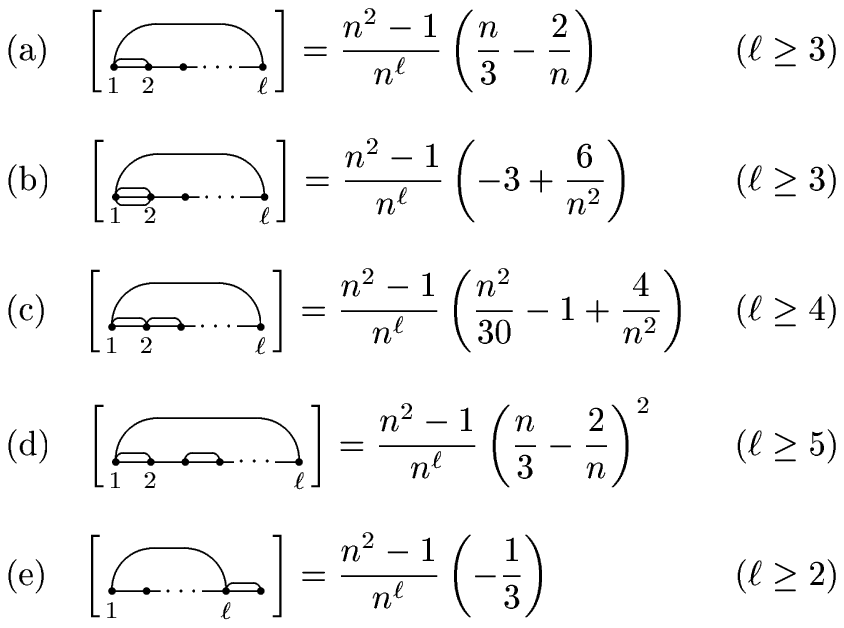}
\caption{Actual calculated values of cumulants.
\label{fig:bondcums}}
\end{figure}

Examples of higher order results are shown
in Fig.~\ref{fig:bondcums}.
We can find simple properties in these cumulants.
\begin{enumerate}
 \item[(i)] A cumulant is reduced by a factor $1/n$ as $\ell$ increases by one.
 \item[(ii)] A double bond in Fig.~\ref{fig:bondcums}(a)
yields a factor of $(\frac{n}{3}-\frac{2}{n})$,
two distant double bonds in Fig.~\ref{fig:bondcums}(d)
yield square of this factor.
In fact, we can make a more general statement:
When multiple bonds are distant,
we can simply make a product of contributions from each of multiple bonds.
\end{enumerate}
We more rigorously derive property (i) in the next subsection
and property (ii) in the subsection after the next.
After that, we formulate the new method using
those properties.

\subsection{The ``contractible'' property}

\begin{figure}[h]
\includegraphics[width= 3.5cm]{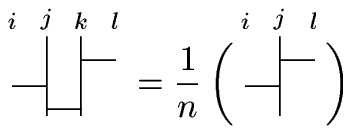}
\caption{ A part of an amida-diagram.
A bond can be deleted to yield a factor $1/n$.
\label{fig:amidacontract}}
\end{figure}

First, we consider a contraction of an amida-diagram.
In an amida-diagram, let us focus on a part
that includes two sites as shown in Fig.~\ref{fig:amidacontract}.
Because of a property of a trace,
the ordering of operators can be cyclically rotated.
In other words,  amida-diagrams have a periodic boundary condition.
Hence
we can put a bond $P_{j,k}$
at the bottom of the amida-diagram
without loss of generality.
Let us compare the l.h.s$.$ and the r.h.s$.$
at each position of bonds.
In the l.h.s., the bottommost bond $P_{j,k}$
unites two cycles at sites $j$ and $k$ of the initial state,
and makes a cycle of the two sites.
At this stage, nothing happens to the r.h.s., and
there is a cycle at $j$.
Then, the rest of the operations of
the l.h.s$.$ and the r.h.s$.$ are the same.
That is, $P_{i,j}$ unites this cycle and another cycle;
when $P_{k,l}$ in the l.h.s$.$ unites cycles or breaks a cycle,
$P_{j,l}$ in the r.h.s$.$ does the same thing.
Therefore,
the number of cycles of the l.h.s$.$ is equal to that of the r.h.s.
As for the denominator of the trace,
the l.h.s$.$ has one more site than the r.h.s.
Hence, in total, the r.h.s$.$ needs a factor $1/n$.
The logic above is true even if site $i$ is the same as site $l$.

\begin{figure}[h]
\includegraphics[width= 4cm]{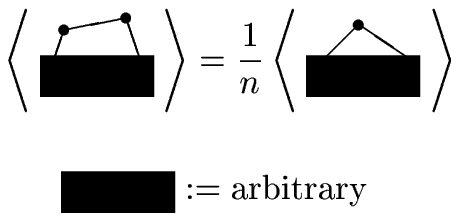}
\caption{
Contraction of a moment bond-diagram.
\label{fig:contractm}}
\end{figure}

The symmetrization of all the operators except $P_{j,k}$ is
equivalent to the symmetrization of all the operators
as explained in Appendix \ref{appn:sym}.
Hence, in all the terms in the symmetrization,
one can put $P_{j,k}$ at the bottom of amida-diagrams
and apply the contraction rule above.
Therefore, we can obtain
a contraction property also for a moment bond-diagram
as shown in Fig.~\ref{fig:contractm}.

\begin{figure}[h]
\includegraphics[width= 7cm]{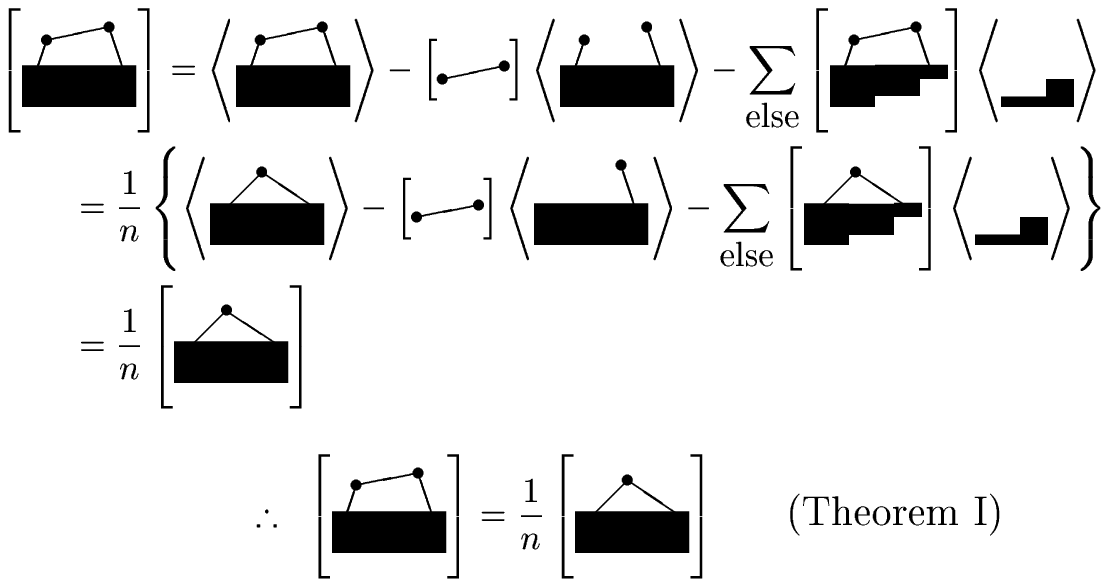}
\caption{Theorem I and its proof.
A solid rectangle represents arbitrary bonds.
In the expansion, the bonds are partitioned into two groups,
which are represented by split rectangles.
\label{fig:contract}}
\end{figure}

Finally, we prove that
a cumulant bond-diagram has the same contraction property
as the corresponding moment bond-diagram does.
We use a mathematical induction for the proof.
\begin{enumerate}
 \item[(i)]
 The simplest nonzero cumulant,
 the l.h.s$.$ of Fig.~\ref{fig:bonddia2}(a), has this contraction property.
 \item[(ii)] Let us think about a certain cumulant bond-diagram.
 We assume that ``the lower-order cumulant bond-diagrams
 have this contraction property.''
 Then, we use the mixed expansion, Eq.~(\ref{eq:mixexpand}),
 pivoting on one of the bonds as shown Fig.~\ref{fig:contract}.
 We use this assumption for the third term.
 The property of moments can be applied to the first and second terms.
\end{enumerate}
Because of (i) and (ii), the contraction property
is true for any cumulant bond diagram that satisfies the condition.
We call this relation the Theorem I.

\subsection{The ``detachable'' property}

\begin{figure}[h]
\includegraphics[width= 8.5cm]{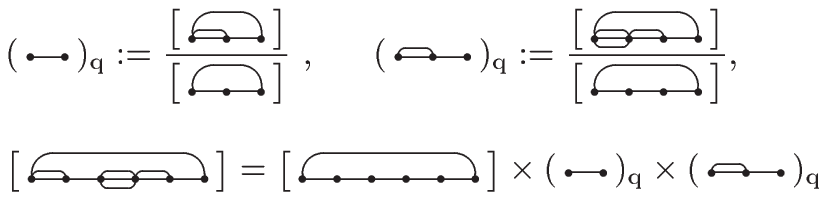}
\caption{A cumulant can be decomposed into
a ``background'' and local quantities.
This property is generally described by Theorem II below.
\label{fig:bonddia3x}}
\end{figure}

Some of the cumulants can be decomposed into
a ``background'' and local quantities.
We give an example in Fig.~\ref{fig:bonddia3x}.

As already appeared several times,
an intersite that has only one bond
plays a special role in this paper.
We call it a single bond here.
On the other hand, an intersite that has more than one bond
is called a multiple bond.

\begin{figure}[h]
\includegraphics[width= 8.5cm]{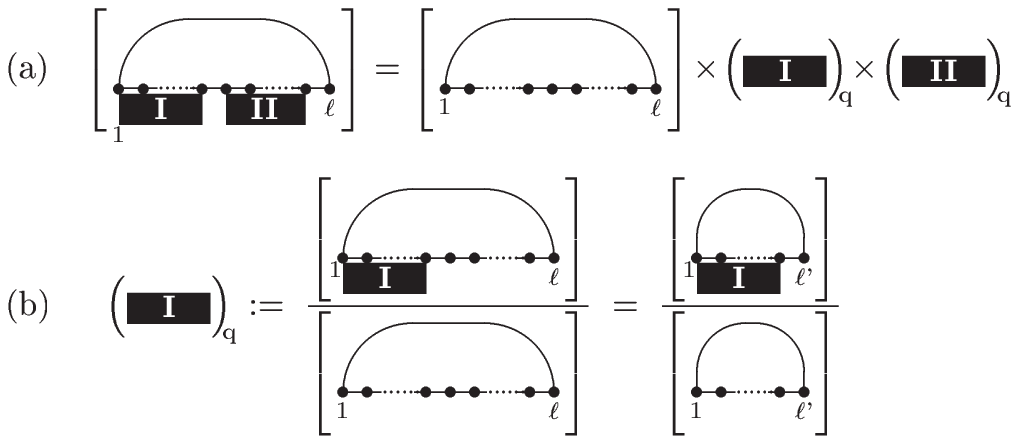}
\caption{Theorem II.
The details are explained in the text.
\label{fig:bonddia3}}
\end{figure}

In this subsection, we prove Theorem II shown in Fig.~\ref{fig:bonddia3}.
Here, a solid rectangle represents arbitrary bonds.
A dotted line stands for a sequence of single bonds.
The bond-diagram in the r.h.s$.$ of (a) has
the same length as that of the l.h.s.;
the bond-diagram in a denominator of (b) has
the same length as that of the numerator.
This decomposition is possible even if
the bond-diagram has more
than two blocks of multiple bonds,
on condition that
the number of single bonds should be
greater than that of blocks.

In other words, the Theorem II is described as follows:
a bond-diagram can be rewritten by
the product of a background and local quantities.
The background is
obtained by replacing each multiple bond with a single bond.
One of the local quantities is obtained by
replacing each multiple bond with a single bond
except those in that block, and dividing it by the background.

\subsubsection{Proof of a lemma}
\label{sec:prooflemma}

\begin{figure}[h]
\includegraphics[width= 7cm]{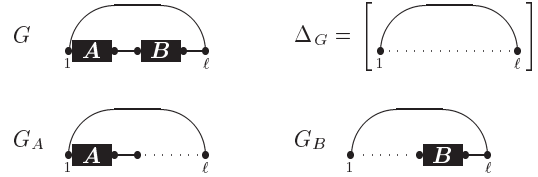}
\caption{Definitions. A dotted line represents
a sequence of single bonds.
The details are explained in the text.
\label{fig:bonddia4}}
\end{figure}

For the proof, we introduce some notation.
It is schematically shown in Fig.~\ref{fig:bonddia4}.
A diagram is denoted by $G$.
Here, solid rectangle $A$ ($B$) represents
``rectangle I (II) and a sequence of single bonds''
in Fig.~\ref{fig:bonddia3} unitedly.
We define $G_A$ ($G_B$)
as a diagram such that each multiple bond in $B$ ($A$) of $G$
is replaced with a single bond.
We define the ``background'' $\Delta_G$
as a cumulant bond-diagram such that
each multiple bond in $A$ and $B$ of $G$
is replaced with a single bond.
Hence, there is a relation,
$\Delta_{G}=\Delta_{G_A}=\Delta_{G_B}$.

\begin{figure}[h]
\includegraphics[width= 7cm]{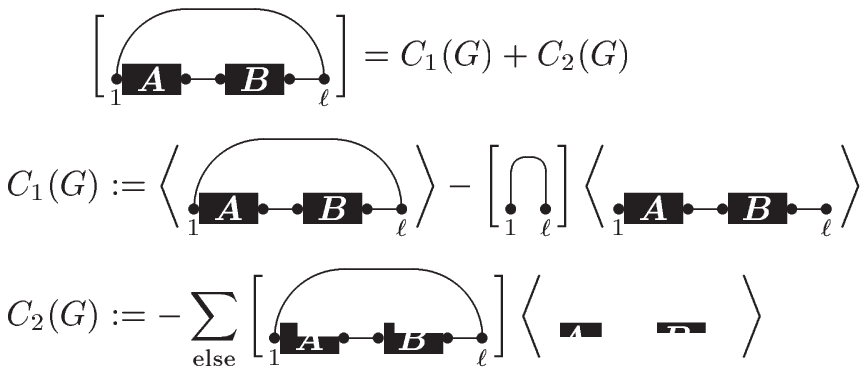}
\caption{
The mixed expansion pivoting on $P_{i,j}$.
The bonds are partitioned into two groups,
which are represented by split rectangles.
\label{fig:bonddia4b}}
\end{figure}

Let us consider the cumulant of $G=P_{i,j}\prod_{k,l} P_{k,l}$.
As is also shown in Fig.~\ref{fig:bonddia4b},
the mixed expansion pivoting on $P_{i,j}$
is written as,
\begin{equation}
[G]_{\rm s} = C_1(G) + C_2(G),
\label{eq:c1plusc2}
\end{equation}
\begin{equation}
C_1(G) :=  \langle G \rangle_{\rm s}
- [P_{i,j}]_{\rm s} \Big\langle \prod_{k,l} P_{k,l}\Big\rangle_{\rm s},
\end{equation}
\begin{equation}
C_2(G) := -\sum_{\rm else}
\Big[P_{i,j} \prod_{k',l'} P_{k',l'}\Big]_{\rm s}
\Big\langle \prod_{k'',l''} P_{k'',l''}\Big\rangle_{\rm s}.
\end{equation}
That is,  the second term of $C_1$ is
decomposition into $P_{i,j}$ and the rest.
The other decompositions are included in $C_2$.
In $C_2$, each cumulant includes more than one exchange operators,
and thus, to give a nonzero cumulant,
each site index has to appear at least
twice as explained in Fig.~\ref{fig:bonddia1b}(c) and (d).
In other words,
each cumulant bond-diagram in $C_2$ has to have a loop.

In order to prove the Theorem II, we first prove a lemma,
\begin{equation}
\frac{C_1(G)}{\Delta_G}
=\frac{C_1(G_A)}{\Delta_G}
\frac{C_1(G_B)}{\Delta_G},
\label{eq:lemma}
\end{equation}
in the following.

\begin{figure}[h]
\includegraphics[width= 7cm]{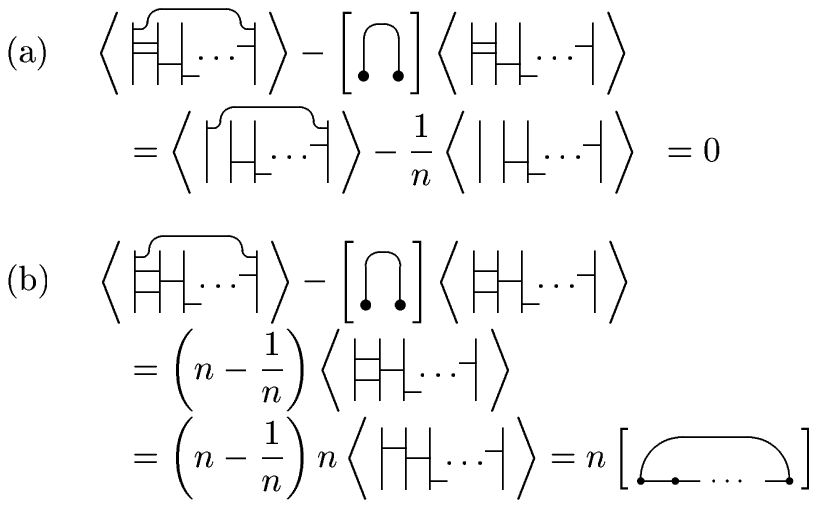}
\caption{
\label{fig:upsilon0}}
\end{figure}

First, we give an explicit example of $C_1$
using the cumulant in Fig.~\ref{fig:bonddia2}(b).
In Sec.~\ref{sec:amida-diagrams}, we have
symmetrized each of the terms in the mixed expansion individually.
Here, however,
we postpone the symmetrization procedure.
That is, as shown in Fig.~\ref{fig:upsilon0},
we first sum amida-diagrams of the terms in $C_1$
with a certain configuration of bonds.
After that, we symmetrize by averaging over all possible
configuration of bonds.
We carry out the symmetrization of all the operators except $P_{1,\ell}$
----- this is equivalent to the symmetrization of all the operators
as shown in Appendix \ref{appn:sym}.
Let us consider Fig.~\ref{fig:upsilon0}(a).
In the amida-diagram in the second term,
sites 1 and $\ell$ belong to different cycles.
Hence, when one more bond $P_{1,\ell}$ is added
(to make the amida-diagram of the first term),
the two cycles are united into one.
Thus, the ratio of these amida-diagrams is $1/n$,
which is equal to $[P_{1,\ell}]_{\rm s}$
in the second term.
Accordingly, these terms cancel out each other.
Therefore,
there is a nonzero contribution to $C_1$ only when
two sites 1 and $\ell$ belong to one cycle
in the ``amida-diagram without $P_{1,\ell}$''
in the second term of $C_1$.
Hereafter, such a configuration is called
a $C_1$-finite configuration.
Figure \ref{fig:upsilon0}(b) is a $C_1$-finite configuration.
Its contribution to $C_1(G_{\rm Fig.\ref{fig:bonddia2}(b)})$
is determined by ``deviation from the background''.
There is an ``extra'' bond compared to the background, and
here it increases the number of cycles by one to make
a factor $n$.
On the other hand,
the ``probability'' for this configuration to occur is 1/3
as shown in Eqs.~(\ref{eq:onethird0}) and (\ref{eq:onethird}).
Therefore we obtain
$C_1(G_{\rm Fig.\ref{fig:bonddia2}(b)})= \frac13 n \,
\Delta_{G_{\rm Fig.\ref{fig:bonddia2}(b)}} $.

The example above is a simple case.
In general,
``the probability of the configuration'' times
``the deviation from the background''
is summed over all the $C_1$-finite configurations,
namely,
\begin{equation}
C_1(G)=\Delta_G
\sum_{\stackrel{\mbox{\tiny $C_1$-finite}}
{\rm \scriptscriptstyle config.}}
({\rm probability})\times({\rm deviation}).
\label{eq:non0confc1}
\end{equation}
Let us consider a diagram $G$ shown in Fig.~\ref{fig:bonddia4}.
Then,
in order to put sites 1 and $\ell$ in one cycle
in the ``amida-diagram without $P_{1,\ell}$'' of $G$,
each of $G_A$ and $G_B$
must also have a $C_1$-finite configuration.
Then, the deviation from the background as a whole is given by
(deviation in $A$)$\times$(deviation in $B$).

\begin{figure}[h]
\includegraphics[width= 8cm]{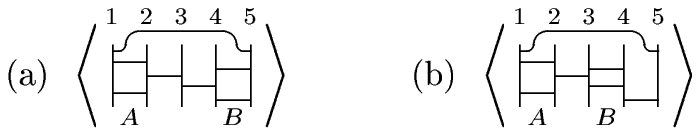}
\caption{
\label{fig:amidaX2}}
\end{figure}

In fact,  ``probability'' is also
a product of those of $A$ and $B$ as shown below.
For example,
the probability of configuration in Fig.~\ref{fig:amidaX2}(a)
is written as
\begin{eqnarray}
&& 2^2
\int_0^{1}  {\rm d}\tau_{12}
\int_{\tau_{12}}^1  {\rm d}\tau_{12}^{\prime}
\int_{\tau_{12}}^{\tau_{12}^{\prime}}  {\rm d}\tau_{23}
\nonumber \\
&& \qquad\quad
\int_0^{1} {\rm d}\tau_{45}
\int_{\tau_{45}}^1 {\rm d}\tau_{45}^{\prime}
\int_{\tau_{45}}^{\tau_{45}^{\prime}}  {\rm d}\tau_{34}.
\label{eq:indepint}
\end{eqnarray}
%
The first three integrals are independent of
the last three integrals, and thus the probability
is written by the product of two parts.
Next, let us think about Fig.~\ref{fig:amidaX2}(b).
If we naively use $\tau$,
the expression of the integrals is complicated.
However, if we utilize the periodic boundary condition of $\tau$
and shift the integration variables as $\rho:=\tau - \tau_{23}$
for $\tau=\tau_{34}, \tau_{34}^{\prime}, \tau_{45}$,
we obtain,
\begin{eqnarray}
&&  2^2
\int_0^{1}  {\rm d}\tau_{12}
\int_{\tau_{12}}^1  {\rm d}\tau_{12}^{\prime}
\int_{\tau_{12}}^{\tau_{12}^{\prime}}  {\rm d}\tau_{23}
\nonumber \\
&& \qquad\quad
\int_0^1  {\rm d}\rho_{34}
\int_{\rho_{34}}^1  {\rm d}\rho_{34}^{\prime}
\int_{\rho_{34}}^{\rho_{34}^{\prime}} {\rm d}\rho_{45}.
\label{eq:indepint2}
\end{eqnarray}
This expression is equivalent to Eq.~(\ref{eq:indepint}).
Hence the probability
is written by the product of two parts again.
The important point in deriving Eq.~(\ref{eq:indepint2}) is
existence of $P_{4,5}$.
Suppose the diagram does not have $P_{4,5}$,
and has $P_{1,4}$ instead of $P_{1,5}$.
The uppermost bond $P_{1,4}$ is always at $\tau=1$.
If the integration variables are shifted,
$1-\tau_{23}$ appears in lower or upper bounds of the integrals.
That is,
the integrals are not written by a product of two parts any more
in that case.

The shift of  the integration variables mentioned above
can generally be applied to any blocks $A$ and $B$.
Hence,  the probability in Eq.~(\ref{eq:non0confc1}) is also
a product of those of $A$ and $B$.
The summation over $C_1$-finite configuration is
double summation over $C_1$-finite configuration of $A$ and
over $C_1$-finite configuration of $B$.
Therefore,
\begin{eqnarray}
\frac{C_1(G)}{\Delta_G}&=&
\sum_{\stackrel{\mbox{\tiny $C_1$-finite}}{{\rm \scriptscriptstyle config.}A}}
({\rm probability} A)({\rm deviation} A)
\nonumber \\ &&
\times
\sum_{\stackrel{\mbox{\tiny $C_1$-finite}}
{{\rm \scriptscriptstyle config.}B}}
({\rm probability} B)({\rm deviation} B)
\nonumber \\
&=& \frac{C_1(G_A)}{\Delta_{G_A}} \frac{C_1(G_B)}{\Delta_{G_B}}.
\end{eqnarray}
By using $\Delta_{G}=\Delta_{G_A}=\Delta_{G_B}$, we obtain the lemma,
Eq.~(\ref{eq:lemma}).

Note that the shift of the integration variables mentioned above
can be applied even if a diagram has more than two blocks
of multiple bonds.
A block and a single bond is regarded as one group
whose members have a common shift of the integration variables.
This is the reason for the restriction of the Theorem II;
the number of single bonds should be
greater than that of blocks.

\subsubsection{Proof of the Theorem II}

In order to focus on the ``multiple bond'' region,
we introduce another notation for moments and cumulants.
Let us think about $G_A$.
A set of bonds, $A$, is partitioned into two subsets
in calculating $C_2(G_A)$.
Let $a$ denote a subset of $A$.
Then, $Q(a)$ denotes the cumulant in which $A$ is replaced by $a$.
In choosing $a$,  at least one bond per intersite must be chosen
to make $Q(a)$ nonzero.
A moment for bonds $A-a$ is denoted by ${\cal M}(A-a)$.
Then, Eq.~(\ref{eq:c1plusc2}) for $G_A$
can be rewritten as
\begin{eqnarray}
Q(A) = C_1(G_A) - \sum_{a\subsetneqq A} Q(a)\,{\cal M}(A-a).
\end{eqnarray}
The l.h.s$.$ is equal to
the excluded term $a=A$ in the summation in the r.h.s$.$
because ${\cal M}(\emptyset) =\langle 1 \rangle_0 =1$.
Therefore, we obtain
\begin{eqnarray}
C_1(G_A) = \sum_{a \subseteqq A } Q(a)\, {\cal M}(A-a).
\label{eq:therefore}
\end{eqnarray}
The corresponding equation for graph $G$ is
\begin{eqnarray}
\frac{C_1(G)}{\Delta}  = \sum_{a\subseteqq A}\sum_{b\subseteqq B}
\frac{Q(a+b)}{\Delta} {\cal M}(A-a+B-b).
\label{eq:Upsilon1-1}
\end{eqnarray}
Since $A$ is not connected to $B$,
\begin{equation}
{\cal M}(A-a+B-b)={\cal M}(A-a){\cal M}(B-b).
\label{eq:momentdecompose}
\end{equation}
On the other hand, according to the lemma and Eq.~(\ref{eq:therefore}),
\begin{eqnarray}
\frac{C_1(G)}{\Delta}  &=&
\frac{C_1(G_A)}{\Delta}
\frac{C_1(G_B)}{\Delta} \nonumber \\
&=& \sum_{a\subseteqq A}\sum_{b\subseteqq B}
\frac{Q(a)}{\Delta} \frac{Q(b)}{\Delta} {\cal M}(A-a){\cal M}(B-b).
\qquad
\label{eq:Upsilon1-2}
\end{eqnarray}
Here, we use a mathematical induction again for the proof.
\begin{enumerate}
 \item[(i)] The Theorem II is true
when there are two extra bonds on the background
as explicitly calculated in Fig.~\ref{fig:bondcums}(d).
 \item[(ii)] We assume that the Theorem II is true
 when the number of extra bonds on the background
is less than that of $G$.
According to the assumption,
\begin{equation}
\frac{Q(a+b)}{\Delta}= \frac{Q(a)}{\Delta} \frac{Q(b)}{\Delta},
\quad (a+b \subsetneqq A+B).
\label{eq:assumption}
\end{equation}
Combine the r.h.s$.$ of
both Eqs.~(\ref{eq:Upsilon1-1}) and (\ref{eq:Upsilon1-2}),
use Eqs.~(\ref{eq:momentdecompose}) and (\ref{eq:assumption}).
Then, all the terms except for $(a,b)=(A,B)$ cancel out and we obtain
\begin{equation}
\frac{Q(A+B)}{\Delta}= \frac{Q(A)}{\Delta} \frac{Q(B)}{\Delta}.
\end{equation}
\end{enumerate}
According to (i) and (ii), the Theorem II is true
for any $G$ that satisfies the conditions.

The proof mentioned above can be applied
even if block $A$ or $B$ includes single bonds.
Then,
the Theorem II for a diagram with two blocks
is repeatedly used to prove
the Theorem II for a diagram with more than two blocks.
Therefore, also a diagram with more than two blocks
has the ``detachable'' property as mentioned first.

\begin{figure}[h]
\includegraphics[width= 8.5cm]{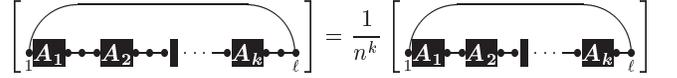}
\caption{ Theorem I'.
\label{fig:Ip}}
\end{figure}

\subsubsection{Theorem I'}

Now that the Theorem II is proved,
the contraction can be written in a more general form.
Although the Theorem I is for a sequence of three single bonds,
sequences of two single bonds also can be contracted
as shown in Fig.~\ref{fig:Ip} as Theorem I',
on condition that
each block must be separated by at least one single bond
and two of the blocks must be separated by a sequence of at least two
single bonds.
The Theorem I' is easily proved using Fig.~\ref{fig:bonddia3}(a);
contract $\Delta_G$ in the r.h.s$.$
and reversely use the same relation to the contracted diagram.

\subsection{The quasi-cumulant method}
Apart from the restriction,
the Theorem II is equivalent to a property of moments
-----
a disconnected moment is equal to a product of connected moments.
In order to effectively use the Theorem II, we define
{\it quasi-moments}
and corresponding {\it quasi-cumulants} below.
Our definition of a quasi-moment
is a factor times a ``local quantity'' in the Theorem II.
Although another definition can be the local quantity itself,
it is less convenient for our purpose
as discussed in Appendix \ref{sec:otherdef}.

\begin{figure}[h]
\includegraphics[width= 4.5cm]{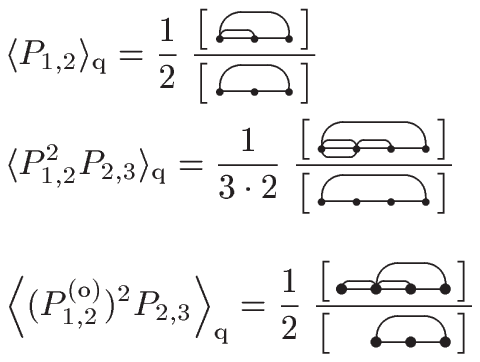}
\caption{Examples of quasi-moments.
\label{fig:figqua}}
\end{figure}

Let us think about the one-dimensional system again.
We define the generating function of quasi-moments as
\begin{eqnarray}
{\cal G}_{\rm qm}^{(i,j)}&:=&
\frac{1}{\lambda_{i,i+1}\lambda_{i+1,i+2}\cdots \lambda_{j-1,j}}
\nonumber \\
&& \times \frac{\langle p_{i,j} \rangle}
{[P_{i,j} \, P_{i,i+1} P_{i+1,i+2} \cdots P_{j-1,j}]_{\rm s}}.
\label{eq:defgqm}
\end{eqnarray}
%
Only up to a finite order, the Theorem II can be applied and
${\cal G}_{\rm qm}$ works as a generating function.
The larger $|i-j|$ is, the higher order the Theorem II is valid up to.
Here, we assume that $|i-j|$ is large enough, and
forget about this restriction
for the time being. It will be commented on later.
Then, the quasi-moments are defined as
\begin{equation}
\langle P_{k,l} \cdots
 P_{y,z} \rangle_{\rm q}:=
\left.
\frac{\partial}{\partial \lambda_{k,l}}
\cdots
\frac{\partial}{\partial \lambda_{y,z}}
 {\cal G}_{\rm qm}^{(i,j)} \right|_{\lambda=0}.
\end{equation}
The zeroth-order quasi-moment satisfies $\langle 1 \rangle_{\rm q}=1$.
Some examples of quasi-moments
are shown in Fig.\ref{fig:figqua}.
More explicitly, the quasi-moments are obtained by
\begin{widetext}
\begin{eqnarray}
\left\langle
 P_{1,2}^{m_1} P_{2,3}^{m_2}\cdots P_{k-1,k}^{m_{k-1}}
\right\rangle_{\rm q}=
\frac{
\left[
 P_{1,2}^{m_1+1} P_{2,3}^{m_2+1}\cdots P_{k-1,k}^{m_{k-1}+1} P_{k,k+1}
 P_{1,k+1}
\right]_{\rm s}}
{ (m_1+1)(m_2+1)\cdots(m_{k-1}+1) \;\;
\left[
 P_{1,2} P_{2,3} \cdots P_{k-1,k} P_{k,k+1} P_{1,k+1}
\right]_{\rm s}},
\label{eq:qmobtain}
\end{eqnarray}
where we have assumed that all the bonds are on the ``background''.
We need a special care
when $P_{k,l}$ is outside the background, namely, $k<i$ or $j<l$ in the
case of ${\cal G}_{\rm qm}^{(i,j)}$.
Such a operator is explicitly denoted by $P_{k,l}^{\rm (o)}$.
Then, we can more generally write as,
\begin{equation}
\left\langle
(P_{1,2}^{\rm (o)})^{m_1} \cdots (P_{i-1,i}^{\rm (o)})^{m_{i-1}}
P_{i,i+1}^{m_i} \cdots P_{k-1,k}^{m_{k-1}}
\right\rangle_{\rm q}=
\frac{
\left[
(P_{1,2}^{\rm (o)})^{m_1} \cdots (P_{i-1,i}^{\rm (o)})^{m_{i-1}}
P_{i,i+1}^{m_i+1} \cdots P_{k-1,k}^{m_{k-1}+1}  P_{k,k+1} P_{i,k+1}
\right]_{\rm s}}
{ (m_i+1) \cdots(m_{k-1}+1)\;\;
\left[
P_{i,i+1} \cdots P_{k-1,k} P_{k,k+1} P_{i,k+1}
\right]_{\rm s}}.
\label{eq:qmobtain2}
\end{equation}
\end{widetext}

Next,
using quasi-moment defined above,
the generating function of quasi-cumulants is defined as,
\begin{equation}
{\cal G}_{\rm qc}^{(i,j)}:= \ln {\cal G}_{\rm qm}^{(i,j)}.
\label{eq:gqcgqm}
\end{equation}
Then, the quasi-cumulants are defined by,
\begin{equation}
[ P_{k,l} \cdots
 P_{y,z} ]_{\rm q}:=
\left.
\frac{\partial}{\partial \lambda_{k,l}}
\cdots
\frac{\partial}{\partial \lambda_{y,z}}
 {\cal G}_{\rm qc}^{(i,j)} \right|_{\lambda=0}.
\label{eq:defqcum}
\end{equation}
According to the Theorem II,
when none of $k$ or $l$ is equal to $y$ or $z$,
\begin{eqnarray}
&& \left\langle
\exp\left[ \sum_{k,l} \lambda_{k,l} P_{k,l} +
      \sum_{y,z} \lambda_{y,z} P_{y,z}   \right]
\right\rangle_{\rm q}
= \nonumber \\ &&
\left\langle \exp\left[\sum_{k,l}\lambda_{k,l} P_{k,l}\right]
\right\rangle_{\rm q}
\left\langle \exp\left[\sum_{y,z}\lambda_{y,z} P_{y,z}\right]
\right\rangle_{\rm q},
\label{eq:prodqm}
\end{eqnarray}
and thus, as discussed in Appendix \ref{appn:fdclvbl},
\begin{equation}
 \left[ \prod_{k,l} P_{k,l}
\prod_{y,z} P_{y,z} \right]_{\rm q}
=0.
\end{equation}
Therefore, only ``connected'' quasi-cumulants remain nonzero.
If only the quasi-moments are used,
we have to classify connected and disconnected quasi-moments.
However, if the quasi-cumulants are used, we need
only {\it connected} quasi-cumulants and can avoid
the complicated classification.

Suppose we set $\lambda_{k,l}=\lambda$ for every $(k,l)$ pair.
Then, ${\cal G}_{\rm qc}^{(i,i+x)}$ $(x>0)$ is
a function of $x$ and $\lambda$.
The quasi-cumulants have a translational symmetry
in the background. For example,
$[P_{i,i+1}^{m}]_{\rm q}=[P_{i+1,i+2}^{m}]_{\rm q}=\cdots=
[P_{i+x-1,i+x}^{m}]_{\rm q}$; and thus,
these ``length-1'' quasi-cumulants can be summed up to
$x\, [P_{i,i+1}^{m}]_{\rm q}$ in ${\cal G}_{\rm qc}^{(i,i+x)}$.
If we in the same way simplify ${\cal G}_{\rm qc}^{(i,i+x)}$,
length-2 quasi-cumulants are multiplied by $x-1$,
length-3 quasi-cumulants are multiplied by $x-2$,
and so on.
The quasi-cumulants including $P_{k,l}^{\rm (o)}$
are located at both ends, and thus, they are multiplied by two.
Namely,
\begin{equation}
{\cal G}_{\rm qc}^{(i,i+x)}=
{\cal G}_{m}^{(x)}+ {\cal G}_{m}^{\rm end}
+O(\lambda^{m+1}),
\label{eq:gqcxend}
\end{equation}
\begin{eqnarray}
{\cal G}_{m}^{(x)}
&:=&\quad x  \sum_{1\le m1 \le m}
\frac{\lambda^{m_1}}{m_1 !} [P_{i,i+1}^{m_1}]_{\rm q}
\nonumber \\
&& + (x-1) \sum_{\stackrel{\scriptstyle 1\le m_1,m_2}{m_1+m_2\le m}}
\frac{\lambda^{m_1+m_2}}{m_1!m_2!}
[P_{i,i+1}^{m_1}P_{i+1,i+2}^{m_2}]_{\rm q} \nonumber \\
&&+\quad \cdots\cdots\cdots \nonumber \\
&&+ (x-m+1) \lambda^m [P_{i,i+1} \cdots P_{i+m-1,i+m}]_{\rm q},
\end{eqnarray}
\begin{eqnarray}
{\cal G}_{m}^{\rm end} \!\!
&:=&\quad 2  \sum_{2 \le m_1 \le m}
\frac{\lambda^{m_1}}{m_1 !} [(P_{i-1,i}^{\rm (o)})^{m_1}]_{\rm q}
\nonumber \\
&& + 2
\sum_{\stackrel{\scriptstyle 2\le m_1, 1\le m_2}{m_1+m_2\le m}}
\frac{\lambda^{m_1+m_2}}{m_1!m_2!}
[(P_{i-1,i}^{\rm (o)})^{m_1} P_{i,i+1}^{m_2}]_{\rm q} \nonumber \\
&& + 2
\sum_{\stackrel{\scriptstyle 2\le m_1, m_2}{m_1+m_2\le m}}
\frac{\lambda^{m_1+m_2}}{m_1!m_2!}
[(P_{i-2,i-1}^{\rm (o)})^{m_1}
(P_{i-1,i}^{\rm (o)})^{m_2}]_{\rm q} \nonumber \\
&&+\quad \cdots\cdots\cdots .
\label{eq:gqcxend2}
\end{eqnarray}
Hence, this expression for arbitrary $x$ can be derived
if we obtain all the quasi-cumulants up to $m$-th order,
which are calculable in small systems.
Therefore, we can obtain $\langle p_{i,i+x} \rangle$
up to $(x+m)$-th order for {\it arbitrary} $x$
by tracing back
Eqs.~(\ref{eq:gqcxend})-(\ref{eq:gqcxend2}),
(\ref{eq:gqcgqm}) and (\ref{eq:defgqm}).
We call this method the quasi-cumulant method (QCM).

The QCM has been done by a brute-force program of Mathematica.
The quasi-cumulants can be calculated by Eq.~(\ref{eq:defqcum}),
that is, differentiating the series of their generating function
with respect to corresponding $J_{i,i+1}$,
where $J_{i,i+1}$ are coupling constants of
the Hamiltonian of a rather general form,
$\sum_i J_{i,i+1} P_{i,i+1}$.
We take the system as small as possible on condition that
it be large enough to correctly calculate those quasi-cumulants.
That is,  quasi-cumulants of length $l$ can be obtained in the system
of length $l+1$.
First, we calculate the series of $\langle p_{i,j} \rangle$
to obtain ${\cal G}_{\rm qm}^{(i,j)}$ up to $m$-th order.
At each order of calculating ${\cal H}^k$,
we exclude terms irrelevant to the derivative
in order to save computational time and memory.
Let us rewrite ${\cal G}_{\rm qm}^{(i,j)}$ as
\begin{equation}
{\cal G}_{\rm qm}^{(i,j)}=
 \sum_{k=0}^m \frac{(-\beta)^k}{k!}
 \left\langle {\cal H}^{k}\right\rangle_{\rm q}
+ O(\beta^{m+1}),
\end{equation}
then, Eq.~(\ref{eq:mixexpand}),
the mixed expansion pivoting on one of $\cal H$ can be used,
\begin{equation}
[{\cal H}^k]_{\rm q}=\langle{\cal H}^k\rangle_{\rm q}
- \sum_{l=1}^{k-1}\frac{(k-1)!}{(l-1)!(k-l)!}
 [ {\cal H}^l  ]_{\rm q}
 \langle {\cal H}^{k-l}  \rangle_{\rm q},
\end{equation}
\begin{equation}
{\cal G}_{\rm qc}^{(i,j)}=
  \sum_{k=1}^m \frac{(-\beta)^k}{k!}
 \left[ {\cal H}^{k}\right]_{\rm q}
+ O(\beta^{m+1}).
\end{equation}
Now we are ready for the derivatives.
The point is that we need at most $m+2$ sites here,
while we can reconstruct the generating function of larger systems
from the quasi-cumulants and thus we can obtain
$\langle p_{i,i+x} \rangle$ up to $(x+m)$-th order arbitrary $x$.

\begin{figure}[h]
\includegraphics[width= 3.5cm]{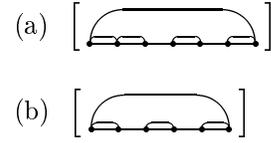}
\caption{The Theorem II is not valid for these diagrams.
\label{fig:notvalid}}
\end{figure}

Finally, we comment on the restriction of the QCM.
The QCM is based on the Theorem II.
Let us consider diagrams shown in Fig.~\ref{fig:notvalid}.
We cannot apply the Theorem II to these diagrams,
and thus Eq.~(\ref{eq:prodqm}) is not valid.
However, we can apply the Theorem II
if the order is lower than these.
Hence, we can use Eqs.~(\ref{eq:gqcxend})-(\ref{eq:gqcxend2}),
only for $m\le x/2$ ($x$: even) or $m\le (x-1)/2$ ($x$: odd).

\begin{figure}[h]
\includegraphics[width= 6cm]{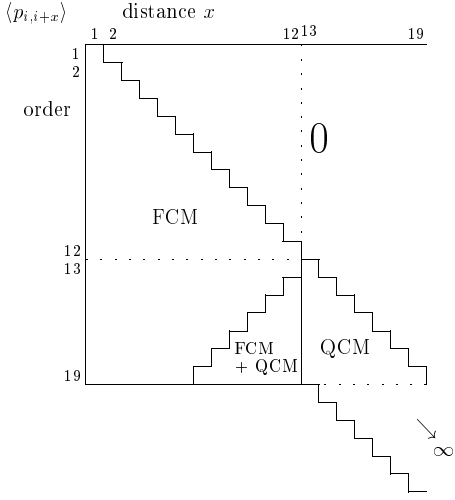}
\caption{We obtained high-temperature series coefficients
of $\langle p_{i,i+x} \rangle$ using the FCM and the QCM complementarily.
The figure shows which method was used for each series coefficient.
Though we have used only all the coefficients up to $O[(\beta J)^{19}]$,
the QCM results are valid for arbitrary $x\ge 13$.
\label{fig:combi}}
\end{figure}

Note that the QCM is useful for contributions from large clusters
rather than those from small clusters.
Hence, by combining the QCM with the FCM,
we can calculate the high-temperature series coefficients
up to a high order.
Figure \ref{fig:combi} shows
which of the FCM and the QCM was used
to obtain the series coefficients
in Ref.~\onlinecite{FK1d}.
We calculated $\langle p_{i,i+x} \rangle$ up to $(x+6)$-th order
for $x\ge 13$ by the QCM.
Note that the QCM results are valid for arbitrary $x\ge 13$,
including the $x \rightarrow \infty$ limit as indicated
in Fig.~\ref{fig:combi}.
The reason why we stopped the QCM at this order was
mainly due to the FCM.
If the FCM can treat larger systems,
the restriction of the Theorem II is less important,
and the QCM can go further.
Our FCM program was for general SU($n$);
we expect that more terms can be obtained
if the interest is restricted to SU(2) in making the FCM program.

In Ref.~\onlinecite{FK1d},
the largest cluster we used for the FCM was composed of $\ell=13$ sites.
As noted in the context of Eq.~(\ref{eq:totcont}),
$\langle {p}_{i,i+x}\rangle'_{\le\ell}$
is correct only up to $O[(\beta J)^{ 2 \ell -x -1}]$,
and higher orders include contributions from the larger clusters.
Hence, for high-order coefficients of $x\le 12$,
corrections to the FCM results must be calculated.
The QCM can be used also for this purpose
because these corrections are a subset of the terms
obtained by the QCM.
We put a fictitious coefficient
on each quasi-cumulant located at both ends in
Eq.~(\ref{eq:gqcxend2}).
These fictitious coefficients carry
information about the cluster-size of a cumulant,
and we can distinguish terms needed for
the corrections to the FCM result.
Coefficients obtained by this procedure
in Ref.~\onlinecite{FK1d}
are indicated by ``FCM+QCM'' in Fig.~\ref{fig:combi}.

Since these series coefficients are obtained as explicit functions of $n$,
the data of the series coefficients are too numerous to
be fully listed in this paper.
Hence, we list here only the results of the QCM,
$\langle p_{i,i+x} \rangle$ up to $(x+6)$-th order,
and the other data will be provided on request.
The coefficients of the series
\begin{equation}
\langle p_{i,i+x} \rangle=
n^{-1 - x}\, \left( -1 + n^2 \right)\,
\sum_m \pi_{x,m}\, (-\beta J)^m
\end{equation}
are given in the following:
\begin{equation}
\pi_{x,x} = 1  \qquad (x \ge 1),
\end{equation}
\begin{equation}
\pi_{x,x+1} =  \frac{\left( -6 + n^2 \right) \,x}{6\,n}
\qquad (x \ge 2),
\end{equation}
\begin{widetext}
\begin{eqnarray}
\pi_{x,x+2} =
\frac{n^{-2}}{360} \,
    \left[ 5\,x^2\,{\left( -6 + n^2 \right) }^2
    - 6\,x\,\left( -30 + 15\,n^2 + 2\,n^4 \right)
    + n^2\,\left( -150 + 7\,n^2 \right)
\right]
&& \nonumber \\
\qquad (x \ge 4),
\end{eqnarray}
\begin{eqnarray}
\pi_{x,x+3} & = & \frac{n^{-3}}{45360}  \,
    \Big[ 35\,x^3\,{\left( -6 + n^2 \right) }^3
  -  126\,x^2\,\left( 180 - 120\,n^2 + 3\,n^4 +   2\,n^6 \right)
 \nonumber \\ && \qquad
  +  2\,x\,\left( -7560 + 17010\,n^2 - 2961\,n^4 + 302\,n^6 \right)
 \nonumber \\ && \qquad
  +  24\,n^2\,\left( 945 + 21\,n^2 - 20\,n^4 \right) \Big]
\qquad  (x \ge 6),
\end{eqnarray}
\begin{eqnarray}
\pi_{x,x+4} & = & \frac{n^{-4}}{5443200}   \,
    \Big[ 175\,x^4\,{\left( -6 + n^2 \right) }^4 -
      1260\,x^3\,{\left( -6 + n^2 \right) }^2\,
       \left( -30 + 15\,n^2 + 2\,n^4 \right)
 \nonumber \\ && \qquad
 +  2\,x^2\,\left( 1247400 - 1965600\,n^2 +
         519750\,n^4 - 48210\,n^6 + 6817\,n^8 \right)
 \nonumber \\ && \qquad
- 6\,x\,\left( -226800 + 1020600\,n^2 - 297990\,n^4 -
             36795\,n^6 + 5468\,n^8 \right)
 \nonumber \\ && \qquad
+ 9\,n^2\,\left( -352800 + 212520\,n^2 - 14850\,n^4 + 3287\,n^6 \right)
      \Big]
\qquad  (x \ge 8),
\end{eqnarray}
\begin{eqnarray}
\pi_{x,x+5}  &=& \frac{n^{-5}}{359251200} \,
    \Big[ 385\,x^5\,{\left( -6 + n^2 \right) }^5
 -  4620\,x^4\,{\left( -6 + n^2 \right) }^3\,
       \left( -30 + 15\,n^2 + 2\,n^4 \right)
 \nonumber \\ &&
 +  44\,x^3\,\left( -2381400 + 3572100\,n^2 -
         1260630\,n^4 + 147915\,n^6 - 16461\,n^8 +  2021\,n^{10} \right)
 \nonumber \\ &&
 -  66\,x^2\,\left( 2268000 - 6350400\,n^2 +
         2891700\,n^4 - 101940\,n^6 - 53013\,n^8 +
         6496\,n^{10} \right)
 \nonumber \\ &&
 +  2\,x\,\big( -35925120 + 264448800\,n^2 -
         182037240\,n^4
 \nonumber \\ &&
\qquad + 17223030\,n^6 - 3830673\,n^8 +
         517472\,n^{10} \big)
 \nonumber \\ &&
 -  36\,n^2\,\left( -6652800 + 7983360\,n^2 -
         60720\,n^4 - 165099\,n^6 + 27776\,n^8 \right)
 \Big]
 \nonumber \\ &&  \qquad (x \ge 10),
\end{eqnarray}
\begin{eqnarray}
\pi_{x,x+6} &=& \frac{n^{-6}}{5884534656000} \,
    \Big[ 175175\,x^6\,{\left( -6 + n^2 \right) }^6
 \nonumber \\ &&
   - 3153150\,x^5\,{\left( -6 + n^2 \right) }^4\,
       \left( -30 + 15\,n^2 + 2\,n^4 \right)
 \nonumber \\ &&
   +  5005\,x^4\,{\left( -6 + n^2 \right) }^2\,
       \left( 3855600 - 4725000\,n^2 + 982800\,n^4 +
         18870\,n^6 + 18947\,n^8 \right)
 \nonumber \\ &&
 -   6006\,x^3\,\big( -306180000 + 731430000\,n^2 -
         413229600\,n^4 + 58660200\,n^6
 \nonumber \\ &&
\qquad         + 2763810\,n^8 -
         1239825\,n^{10} + 126908\,n^{12} \big)
 \nonumber \\ &&
 + 13\,x^2\,\big( 172260950400 - 745940210400\,n^2 +
         577975305600\,n^4 - 101282227200\,n^6
 \nonumber \\ &&
\qquad         + 6937664580\,n^8 - 2448983250\,n^{10} +
         265586033\,n^{12} \big)
 \nonumber \\ &&
 -  6\,x\,\big( -163459296000 + 1777619844000\,n^2 -
         2058679022400\,n^4
 \nonumber \\ &&
\qquad       + 396577981800\,n^6
       + 31224022830\,n^8 - 11642353905\,n^{10} +
         1392291386\,n^{12} \big)
 \nonumber \\ &&
   +  135\,n^2\,\big( -32691859200 + 61509127680\,n^2 -
         15679984320\,n^4 + 695398704\,n^6
 \nonumber \\ &&
\qquad   -  446293302\,n^8 + 62451523\,n^{10} \big)
 \Big]
\qquad (x \ge 12).
\end{eqnarray}
\end{widetext}

Even for $\ell\le 12$, the QCM is valid for a number of
series coefficients,
and should coincide with the results of the FCM or the ``FCM+QCM''.
We have used this property for checking whether the computer codes
are correct.

\section{A technique for the specific heat}
\label{sec:newsh}

Our strategy for the specific heat in Ref.~\onlinecite{FK1d}
was similar to that for the correlation functions.
That is, the most time-consuming part of the FCM
is calculated by another method instead.
If only the FCM is used,
the series for $\langle P_{i,i+1} \rangle$ up to $O[(\beta J)^M]$
requires systems with
$\ell\le M/2+1$ ($M$: {\rm even})
or $\ell\le (M+1)/2+1$ ($M$: {\rm odd}).
In Ref.~\onlinecite{FK1d},
we calculated $\langle P_{i,i+1} \rangle$ up to $O[(\beta J)^{22}]$.
However, the FCM was used only for $\ell \le 11$;
contributions from the required largest cluster,
namely,
the lowest- and the second-lowest-order nonzero contribution
from a cluster,
was directly calculated
by a new technique explained below.

In fact, we can use
$\langle \cdots \rangle_0$ as a moment and
define a cumulant\cite{Fulde}
without the symmetrization.
The generating functions of the moments and the cumulants are defined as
\begin{equation}
{\cal G}_{0\rm m}^{i_1,j_1,\cdots,i_k,j_k}:=
 \left\langle
e^{\lambda_1 P_{i_1,j_1} }
e^{\lambda_2 P_{i_2,j_2} }
\cdots
e^{\lambda_k P_{i_k,j_k} }
\right \rangle_0,
\end{equation}
\begin{equation}
{\cal G}_{0\rm c}^{i_1,j_1,i_2,j_2,\cdots,i_k,j_k}:=
\ln {\cal G}_{0\rm m}^{i_1,j_1,i_2,j_2,\cdots,i_k,j_k},
\end{equation}
respectively. Then, $\langle P_{i_1,j_1} \cdots P_{i_k,j_k}  \rangle_0$
works as a moment, and
the corresponding cumulant is define by
\begin{eqnarray}
[ P_{i_1,j_1} \cdots P_{i_k,j_k} ]_0
:= \left.
\frac{\partial}{\partial \lambda_1}
\cdots
\frac{\partial}{\partial \lambda_k}
{\cal G}_{0\rm c}^{i_1,j_1,\cdots,i_k,j_k}
 \right|_{\lambda=0}.
\end{eqnarray}
Here, the ordering of exchange operators in $[\cdots]_{0}$ is important.
In other words,
the result can be changed by exchanging positions
of uncommutative operators.
Note that simplification such as $P_{1,2}P_{1,2}=1$ is
{\it not} possible in $[\cdots]_0$
while it is possible in $\langle\cdots\rangle_0$.
In expanding the cumulants using the moments,
the ordering of operators in $\langle\cdots\rangle_0$ must be
the same as that in $[\cdots]_0$,
for example,
$[ P_{1,2} P_{2,3} ]_0 =  \langle P_{1,2} P_{2,3} \rangle_0
- \langle P_{1,2} \rangle_0 \langle P_{2,3} \rangle_0$.
This cumulant is represented by the cumulant version of an
amida-diagram.
Hereafter, we distinguish between a moment amida-diagram and
a cumulant amida-diagram by bracketing the diagram by
$\langle\cdots\rangle$ and $[\cdots]$, respectively.

The free energy is rewritten as
\begin{equation}
F = - \frac{1}{\beta} \ln Z_0
 -\frac{1}{\beta} \sum_{m=1}^\infty \frac{(-\beta)^{m}}{m!}
[{\cal H}^m]_{0},
\end{equation}
%
\begin{eqnarray}
[{\cal H}^m]_{0} &=&
\sum_{(i_1,j_1)} \cdots \sum_{(i_{m},j_{m})}
J_{i_1,j_1}\cdots J_{i_{m},j_{m}}
\nonumber \\ &&
\qquad\qquad\qquad \times
   \left[
P_{i_1,j_1} \cdots P_{i_{m},j_{m}}
   \right]_{0}.
\end{eqnarray}
These are equivalent to Eqs.~(\ref{eq:hmcs0}) and (\ref{eq:hmcs})
except for the subscript ``0'' replacing ``s''.
Here, however, the cumulants are generally dependent on the ordering of
the operators while ``s'' makes many terms equivalent.

\begin{figure}[h]
\includegraphics[width= 7cm]{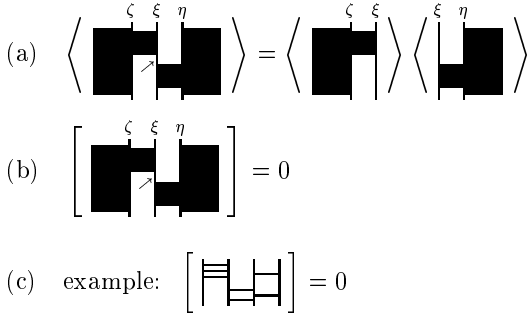}
\caption{
Bonds are located only in the painted regions.
When an amida-diagram can be separated into two diagrams
by cutting only one point on a vertical line (indicated by an arrow),
the left and right blocks are independent of each other.
\label{fig:cumzero}}
\end{figure}

We can use general property of moments and cumulants again.
If,
\begin{eqnarray}
&& \left\langle
\prod_{k} \exp\left[\lambda_k P_{i_k,j_k} \right]
\prod_{l} \exp\left[\lambda_l P_{i_l,j_l} \right]
\right\rangle_0
= \nonumber \\ &&
\left\langle
\prod_{k} \exp\left[\lambda_k P_{i_k,j_k} \right]
\right\rangle_0
\left\langle
\prod_{l} \exp\left[\lambda_l P_{i_l,j_l} \right]
\right\rangle_0,
\end{eqnarray}
then, the derivative of its logarithm gives,
\begin{equation}
\left[
\prod_{k}  P_{i_k,j_k}
\prod_{l}  P_{i_l,j_l}
\right]_0
=0.
\end{equation}
When none of $i_k$ or $j_k$ is equal to $i_l$ or $j_l$,
this relation is obviously satisfied.
Furthermore, the cumulant is equal to zero also when
an amida-diagram is separated into two diagrams
by cutting only one point on a vertical line
as shown in Fig.~\ref{fig:cumzero}.
For example, the cumulant versions of
the amida-diagrams in Fig.~\ref{fig:defdia}(a) and (b)
are equal to zero.
Let us think about the moment amida-diagram, Fig.~\ref{fig:cumzero}(a),
where bonds are located only in the painted regions.
Bonds located left of $\xi$ are denoted by
$P_{i_l,j_l}$, and  those located right of $\xi$
are denoted by $P_{i_r,j_r}$.
The diagram can be written as
$\langle \prod_l P_{i_l,j_l} \prod_r P_{i_r,j_r} \rangle_0$
by exchanging positions of commutative $P_{i_l,j_l}$ and $P_{i_r,j_r}$.
Let us imagine calculating each of
$\prod_l P_{i_l,j_l}$ and $\prod_r P_{i_r,j_r}$
by counting cycles,
and make the product of these two at the end.
This final operation unites
the two cycles relevant to $\xi$ into one cycle
because $\prod_l P_{i_l,j_l}$ and $\prod_r P_{i_r,j_r}$ are
sharing only one site $\xi$.
Hence,
the number of cycles in the l.h.s$.$ is less than that in the r.h.s$.$
``by one''.
However, the number of sites in the l.h.s$.$ is also less
than that in the r.h.s$.$ ``by one''.
As a result, the l.h.s$.$ is equal to the r.h.s.
We can make the same argument to
$\langle \prod_l e^{\lambda_l P_{i_l,j_l}}
\prod_r e^{\lambda_r P_{i_r,j_r}} \rangle_0$.
Therefore, $[ \prod_l P_{i_l,j_l} \prod_r P_{i_r,j_r} ]_0=0$.

\begin{figure}[h]
\includegraphics[width= 8cm]{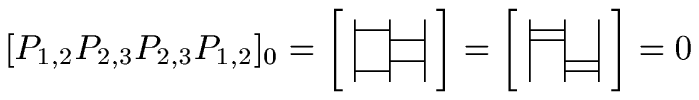}
\caption{
\label{fig:defdia2}}
\end{figure}

Here, we must be careful about the periodic boundary condition
of amida-diagrams.
If a diagram satisfies the condition above
by using the periodic boundary condition,
the amida-diagram is also equal to zero.
We give an example in Fig.~\ref{fig:defdia2}.
At a sight, $[ P_{1,2}P_{2,3}P_{2,3}P_{1,2}]_0$
does not satisfy the condition above.
However, it is equivalent to
$[ P_{1,2} P_{1,2}P_{2,3}P_{2,3}]_0 =0$.

\begin{figure}[h]
\includegraphics[width= 4cm]{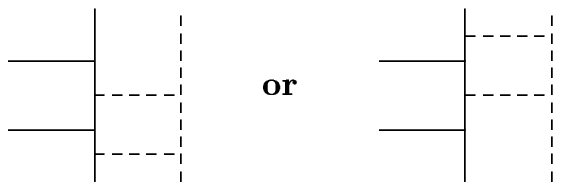}
\caption{
\label{fig:amida1st2}}
\end{figure}

\begin{figure}[h]
\includegraphics[width= 7cm]{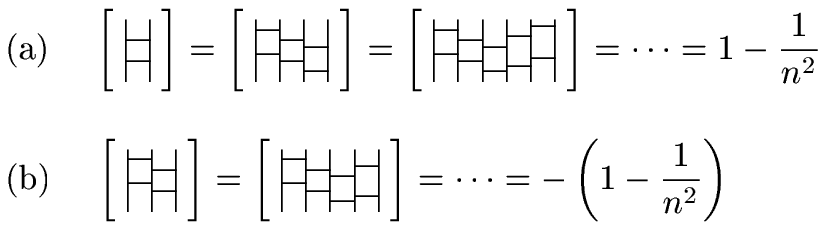}
\caption{
\label{fig:amida1st1}}
\end{figure}

Suppose there are two bonds per nearest-neighbor pair as shown in
Fig.~\ref{fig:amida1st1}.
It is a contribution from cluster-size $\ell$
for $O[(\beta J)^{2(\ell-1)-1}]$ of $\langle P_{i,i+1} \rangle$,
or equivalently,  $O[(\beta J)^{2(\ell-1)}]$ of the free energy.
Because of the property mentioned above,
many configurations of bonds make cumulant amida-diagrams equal to zero.
In order to give a nonzero contribution,
at each vertical line bonds at left and right have to
appear alternately.
We can make such diagrams by repeating a enlarging procedure
shown in Fig.~\ref{fig:amida1st2}.
That is, there are two possibilities for next bonds.
In fact, the value of such a cumulant amida-diagram
does not depend on those configurations as explained
in Appendix \ref{sec:proofspht}.
Hence, we can just multiply the value of the cumulants by
the probability for such nonzero configurations.
We have calculated the probability by Mathematica
using integrals like Eqs.~(\ref{eq:onethird}) and (\ref{eq:indepint}).

\begin{figure}[h]
\includegraphics[width= 8.5cm]{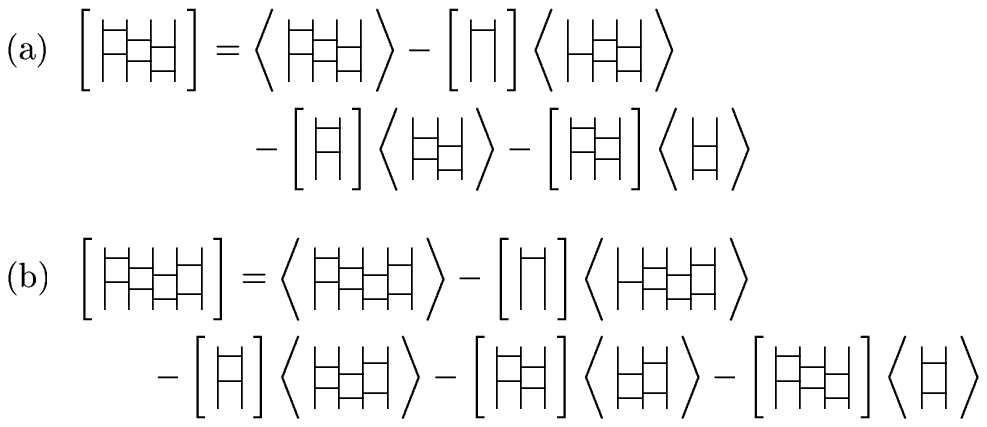}
\caption{
\label{fig:amida1st4}}
\end{figure}

The values of the cumulants are shown in Fig.~\ref{fig:amida1st1}.
Namely, the cumulant is equal to $(1-n^{-2})$ when $\ell$ is an even number,
and $-(1-n^{-2})$ when $\ell$ is an odd number.
We can prove this by using
the mixed expansion pivoting on one of $P_{1,2}$
shown in Fig.~\ref{fig:amida1st4}.
The first term is the moment.
The second term is $[P_{1,2}]$ times the moment of the others.
The rest of the terms are
obtained by separating the diagram at sites from 2 to $\ell-1$.
In fact, only the last term survives and
every term else cancels out one another.
When $\ell$ is an even number,
the sum of the first, second and third terms,
the sum of the fourth and fifth terms, \dots,
are equal to zero.
When $\ell$ is an odd number,
the sum of the first and second terms,
the sum of the third and fourth terms, \dots,
are equal to zero.
The details are explained in Appendix \ref{sec:proofspht}.


\begin{figure}[h]
\includegraphics[width= 8cm]{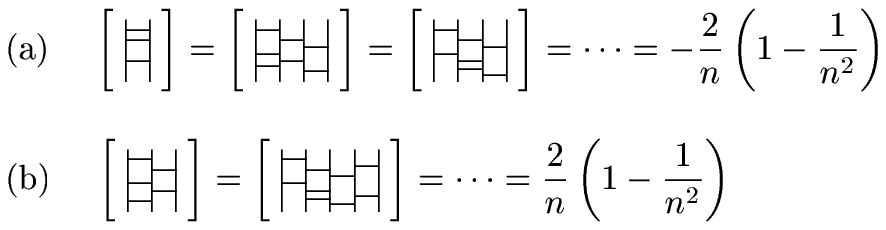}
\caption{
\label{fig:amida2nd}}
\end{figure}

We can calculate also the next order
by the same technique as above.
Now, one of the nearest-neighbor pairs has three bonds.
The value of the cumulant amida-diagram does not depend on
the position of this triple bond
as shown in Fig.~\ref{fig:amida2nd}.
That is,
the cumulant is equal to $-\frac{2}{n}(1-\frac{1}{n^2})$
when $\ell$ is an even number,
and equal to
$\frac{2}{n}(1-\frac{1}{n^2})$ when $\ell$ is an odd number.

\section{summary}
\label{sec:con}

We have formulated a new method of high-temperature series expansion
using the SU($n$) Heisenberg model.
It is designed for efficiently calculating
contributions from large clusters, which is actually
the most time-consuming part of a standard method.
The net contributions from a cluster
to high-temperature series
start from a certain order.
We have focused on this property,
and considered deviation from the
lowest-order nonzero contribution.

\begin{figure}[h]
\includegraphics[width= 1cm]{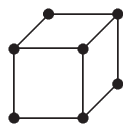}
\caption{An example of clusters unsuitable for our method.
\label{fig:suitable}}
\end{figure}

Although we have mainly used one-dimensional systems
to explain the new method, many of the techniques
can be used also in higher dimensions.
In contrast to one-dimensional systems, however,
the calculation requires various clusters,
and some of the clusters are unsuitable for our method.
Figure \ref{fig:suitable} shows an example.
The main advantage to use our method
is that diagrams can be reduced to smaller diagrams.
However, we cannot apply it to a cluster in Fig.~\ref{fig:suitable}
because it has many sites at which more than two bonds meet,
and chains connecting those sites are too short.
Nevertheless,
the number of embeddings of such unsuitable clusters
in the infinite lattice is small,
and thus contributions from them are usually small.
In fact, the number of embeddings of open chains, the
most preferable clusters for our method, is the largest.
Hence, the dominant contributions can be calculated by our method.

We expect that similar approaches are possible also for
other models, namely, for Hamiltonians of lower symmetries.

\acknowledgments
The author is grateful to M.~Arikawa for
indicating him the equivalence of an amida-lottery and a permutation.
He also thanks N.~Shannon, K.~Penc and Y.~Kuramoto for
useful comments on the manuscript.
This work was financially supported by Visitors' Program of the MPI--PKS
and the Japan Society for the Promotion of Science.

\appendix
\section{relations derived from the isotropy}

\label{app:derive}

For example,
using $n^2-1$ generators such as
\begin{equation}
{\cal X}^{\alpha\beta}:=
\sqrt{\frac{n}{2}}\left( X^{\alpha\beta}+X^{\beta\alpha}\right)
\quad (1\le \alpha < \beta \le n),
\end{equation}
\begin{equation}
{\cal Y}^{\alpha\beta}:=
-i\sqrt{\frac{n}{2}}\left( X^{\alpha\beta}-X^{\beta\alpha}\right)
\quad (1\le \alpha < \beta \le n),
\end{equation}
%
\begin{eqnarray}
D^{(\alpha)}&:=&
\sqrt{\frac{n}{\alpha(\alpha-1)}}
\left(
\sum_{\gamma=1}^{\alpha-1}X^{\gamma\gamma}-(\alpha-1)X^{\alpha\alpha}
\right) \nonumber \\
&& \qquad\qquad\qquad\qquad
 (2\le \alpha \le n),
\end{eqnarray}
%
the exchange operator is rewritten as
%
\begin{eqnarray}
n P_{i,j}
  &=& \sum_{1\le \alpha < \beta \le n}
  \left(
   {\cal X}_i^{\alpha\beta} {\cal X}_j^{\alpha\beta}
   + {\cal Y}_i^{\alpha\beta}   {\cal Y}_j^{\alpha\beta}
   \right) \nonumber \\
 & &\;\;\;\;  +\sum_{2\le \alpha \le n}  D_i^{(\alpha)}  D_j^{(\alpha)}
+1.
\end{eqnarray}
Let us prove that each component contributes equally, namely,
%
\begin{eqnarray}
&&
\langle {\cal X}_i^{\alpha\beta} {\cal X}_j^{\alpha\beta} \rangle
=
\langle {\cal Y}_i^{\alpha\beta} {\cal Y}_j^{\alpha\beta} \rangle
=
\langle D_i^{(\alpha)} D_j^{(\alpha)} \rangle
 \nonumber \\
&& \qquad\qquad
=
\frac{n \langle  P_{i,j} \rangle - 1 }{n^2-1} .
\label{eq:equalcontr}
\end{eqnarray}
%
\begin{enumerate}
 \item[(i)]
The Hamiltonian is invariant under
re-labeling of colors. Therefore,
$\langle {\cal X}_i^{\alpha\beta} {\cal X}_j^{\alpha\beta} \rangle$
does not depend on $\alpha$ or $\beta$.
 \item[(ii)] $
\langle {\cal X}_i^{\alpha\beta} {\cal X}_j^{\alpha\beta} \rangle
=
\langle {\cal Y}_i^{\alpha\beta} {\cal Y}_j^{\alpha\beta} \rangle
$ for arbitrary $\alpha$, $\beta$,
because these ${\cal X}$, ${\cal Y}$ operators are
$x$, $y$ components of Pauli matrix $(\times \sqrt{n/2})$
regarding only $\alpha$ and $\beta$.
 \item[(iii)] We can apply the same argument also for $z$-component,
and thus
$
\langle {\cal X}_i^{\alpha\beta} {\cal X}_j^{\alpha\beta} \rangle
=\langle D_i^{(2)} D_j^{(2)} \rangle
$.
 \item[(iv)] By a direct calculation using
\[
\langle
X_i^{\alpha \alpha} X_j^{\alpha \alpha}
\rangle
=
\langle
X_i^{1 1} X_j^{1 1}
\rangle
\] for arbitrary $\alpha$, and
\[
\langle
X_i^{\alpha \alpha} X_j^{\beta \beta}
\rangle
=
\langle
X_i^{1 1} X_j^{2 2}
\rangle
\] for arbitrary $\alpha\neq\beta$,
we can derive
\begin{eqnarray*}
\langle D_i^{(\alpha)} D_j^{(\alpha)} \rangle
& = &
n \Big(
\langle
X_i^{1 1} X_j^{1 1}
\rangle
-
\langle
X_i^{1 1} X_j^{2 2}
\rangle \Big)   \\
& = & \langle D_i^{(2)} D_j^{(2)} \rangle
\end{eqnarray*}
for arbitrary $\alpha$.
\end{enumerate}
From (i)-(iv), we obtain Eq.~(\ref{eq:equalcontr}).

Using Eq.~(\ref{eq:equalcontr}),
we can derive
\begin{eqnarray}
\langle X_i^{\alpha\beta} X_j^{\beta\alpha} \rangle
&=& \frac{1}{n}
\langle {\cal X}_i^{\alpha\beta} {\cal X}_j^{\alpha\beta} \rangle
\nonumber
\\
&=&
\frac{1}{n^2-1} \left( \langle  P_{i,j} \rangle -\frac{1}{n} \right),
\end{eqnarray}
for $\alpha \neq \beta, i\neq j$.

In addition, for $i= j$,
since the average of $|\alpha\rangle \langle \alpha|$
does not depend on $\alpha$,
\begin{equation}
\langle X_i^{\alpha\beta} X_i^{\beta\alpha} \rangle
=
\Big\langle|\stackrel{i}{\alpha}\rangle \langle \stackrel{i}{ \alpha }|
\Big\rangle
=
\frac{1}{n}\sum_{\gamma=1}^n
\Big\langle|\stackrel{i}{\gamma}\rangle \langle \stackrel{i}{ \gamma }|
\Big\rangle.
\end{equation}
Then, using
\begin{equation}
\sum_{\gamma=1}^n
|\stackrel{i}{\gamma}\rangle \langle \stackrel{i}{ \gamma }|
=1,
\end{equation}
we obtain
\begin{equation}
\langle X_i^{\alpha\beta} X_i^{\beta\alpha} \rangle
=\frac{1}{n} \langle 1 \rangle
=\frac{1}{n}.
\end{equation}

\section{Moments and cumulants for classical variables}
\label{appn:fdclvbl}

In this section, we review
fundamental properties of moments and cumulants of classical variables
$x_i \; (i=1,2,\cdots)$.
Note that some other properties are written in Sec.~\ref{sec:mcclassic}.
Here, the expectation value
with respect to the distribution of $\{x_i\}$
is denoted by $ \langle \cdots \rangle_{x}$.
The generating function of moments is defined as
\begin{equation}
\left\langle e^{\sum_i \lambda_i x_i} \right \rangle_x.
\end{equation}
That is, the moments are derived by
\begin{equation}
\langle x_j x_k \cdots x_l \rangle_x=
\left.
\frac{\partial}{\partial\lambda_j}
\frac{\partial}{\partial\lambda_k}
\cdots
\frac{\partial}{\partial\lambda_l}
\left\langle e^{\sum_i \lambda_i x_i} \right \rangle_x
\right|_{\lambda=0},
\end{equation}
where the subscript $\lambda=0$ represents that
all the $\lambda$-variables are set to zero.
On the other hand, cumulants are denoted by $[\cdots]_x$
and defined as
\begin{equation}
[ x_j x_k \cdots x_l ]_x =
\left.
\frac{\partial}{\partial\lambda_j}
\frac{\partial}{\partial\lambda_k}
\cdots
\frac{\partial}{\partial\lambda_l}
\ln \left\langle e^{\sum_i \lambda_i x_i} \right \rangle_x
\right|_{\lambda=0}.
\end{equation}
Namely,
the generating function of the cumulants is written as
\begin{equation}
\ln \left\langle e^{\sum_i \lambda_i x_i} \right \rangle_x
=\left [ e^{\sum_i \lambda_i x_i} -1 \right ]_{x}.
\end{equation}

A cumulant is equal to zero if the variables in $[\cdots]_x$
can be partitioned into two groups
which are
independent of each other in averaging $\langle \cdots \rangle_x$.
Suppose $\{x_i\}$ is independent of $\{x_j\}$.
Let us in advance set $\lambda_k$ to zero if $x_k$ is not included in
either $\{x_i\}$ or $\{x_j\}$.
Then,
\[
\langle e^{\sum_i \lambda_i x_i + \sum_j \lambda_j x_j} \rangle_x =
\langle e^{\sum_i \lambda_i x_i} \rangle_x
\langle e^{\sum_j \lambda_j x_j} \rangle_x.
\]
By taking the logarithm, we obtain
\[
 \ln \langle e^{\sum_i \lambda_i x_i} \rangle_x
+\ln \langle e^{\sum_j \lambda_j x_j} \rangle_x.
\]
That is, the generating function of the cumulants does not have
any term which has a product of $\lambda_i$ and $\lambda_j$.
Thus, the corresponding cumulant is equal to zero.

There are relations between moments and cumulants.
Moments can be expanded using cumulants,
for example,
\begin{eqnarray*}
 \langle x_1 \rangle_x &=& [ x_1 ]_{x}, \\
 \langle x_1 x_2\rangle_x &=&
[ x_1 x_2]_{x} +[ x_1 ]_{x} [ x_2 ]_{x}, \\
 \langle x_1 x_2 x_3 \rangle_x &=&
[ x_1 x_2 x_3 ]_{x} +
[ x_1 x_2 ]_{x} [ x_3 ]_{x} +
[ x_1 x_3]_{x} [ x_2 ]_{x}  \\ &&+
[ x_1 ]_{x} [ x_2 x_3]_{x} +
[ x_1]_{x} [ x_2]_{x}
[ x_3 ]_{x}.
\end{eqnarray*}
These relations are generally written as
\begin{equation}
\langle x_i  \cdots x_l \rangle_{x}
= \sum_\zeta \sum_{{\cal P}(\xi,\zeta)}
[ x_i\cdots x_j ]_{x}
\cdots
[ x_k \cdots x_m ]_{x}.
\label{eq:m2c}
\end{equation}
These summations include every partition of $x$-variables.
Here, $\xi$ is the number of $x$-variables in the bracket
in the l.h.s.
Then, the summation denoted by ${\cal P}(\xi,\zeta)$
is taken over every partition of $\xi$ elements into
$\zeta$ groups ($1\le \zeta \le \xi$).
In other words, $\zeta$ is the number of brackets $[\cdots]_x$
in the r.h.s.
Here, each $[\cdots]_x$ includes at least one $x$.

Conversely, cumulants can be expanded using moments, for example,
\begin{eqnarray*}
[ x_1 ]_{x} & = & \langle x_1 \rangle_x, \\
{} [ x_1 x_2]_{x}
&=& \langle x_1 x_2\rangle_x
 -\langle x_1 \rangle_{x} \langle x_2 \rangle_{x}, \\
{} [ x_1 x_2 x_3 ]_x &=&
\langle x_1 x_2 x_3 \rangle_{x} -
\langle x_1 x_2 \rangle_{x} \langle x_3 \rangle_{x} -
\langle x_1 x_3\rangle_{x} \langle x_2 \rangle_{x}  \\ &&-
\langle x_1 \rangle_{x} \langle x_2 x_3\rangle_{x} +2
\langle x_1\rangle_{x} \langle x_2\rangle_{x}
\langle x_3 \rangle_{x}.
\end{eqnarray*}
In general, Eq.~(\ref{eq:m2c}) is inverted to
\begin{eqnarray}
[ x_i  \cdots x_\ell ]_{x}
& = & \sum_\zeta \sum_{{\cal P}(\xi,\zeta)} (-1)^{\zeta-1} (\zeta-1)!
\nonumber \\ && \quad \times
\langle x_i\cdots x_j \rangle_{x}
\cdots \langle x_k \cdots x_m \rangle_{x}.
\label{eq:c2m}
\end{eqnarray}

\section{symmetrization}
\label{appn:sym}

Let us consider expanding ${\cal H}^m$.
It includes all the orderings of $P_{i,j}$ operators.
In other words, it is already symmetrized.
Therefore,
\begin{equation}
 \langle {\cal H}^m \rangle_{\rm s}=
 \langle {\cal H}^m \rangle_{0}.
\end{equation}
Accordingly,
\begin{equation}
 \langle e^{- \beta {\cal H}} \rangle_{\rm s}=
 \langle e^{- \beta {\cal H}} \rangle_{0}.
\label{eq:sym0exp}
\end{equation}

Because of a a property of trace, one can
cyclically rotate variables in the average.
In each term of the symmetrization,
let us put a certain operator, for example $P_{i_k,j_k}$,
at the leftmost. Then, $k$ terms give the same ordering.
Namely,
\begin{eqnarray}
&&
\sum_\sigma
\big\langle
P_{i_{\sigma(1)},j_{\sigma(1)}} P_{i_{\sigma(2)},j_{\sigma(2)}}\cdots
 P_{i_{\sigma(k)},j_{\sigma(k)}} \big\rangle_0 =
\nonumber\\
&&  k
\big\langle P_{i_{k},j_{k}}
\sum_{\tilde{\sigma}}
 P_{i_{\tilde{\sigma}(1)},j_{\tilde{\sigma}(1)}}
 P_{i_{\tilde{\sigma}(2)},j_{\tilde{\sigma}(2)}}\cdots
 P_{i_{\tilde{\sigma}(k-1)},j_{\tilde{\sigma}(k-1)}} \big\rangle_0,
\nonumber\\ &&
\end{eqnarray}
where $\tilde{\sigma}$ is a permutation of $k-1$ elements.
The factor $k$ cancels out $k$ in $k!$ in Eq.~(\ref{eq:symmetrization}).
That is, the symmetrization of all the operators
except one is
equivalent to the symmetrization of all the operators.
Therefore, $\langle P_{i,j}{\cal H}^m  \rangle_0$ is also
already symmetrized, and thus
\begin{equation}
 \langle P_{i,j}e^{- \beta {\cal H}} \rangle_{\rm s} =
 \langle P_{i,j}e^{- \beta {\cal H}} \rangle_{0}.
\end{equation}

\section{proofs for cumulants}
\label{app:addprf}

Here, we prove properties of cumulants in Fig.~\ref{fig:bonddia1b}.
For a simplicity, we use the SU(2) notation here.
However, for general SU($n$), one can easily replace spin operators below
by
$ {\cal X}_i^{\alpha\beta}$,
$ {\cal Y}_i^{\alpha\beta}$,
$ D_i^{(\alpha)}$ in Appendix \ref{app:derive}.

In order to consider Fig.~\ref{fig:bonddia1b}(c) and (d).
Let us expand the l.h.s$.$ of Eq.~(\ref{eq:cumequal}) using moments,
namely, in terms of Eq.~(\ref{eq:c2m}).
Then, let us rewrite the $p_{i,j}$ operators using spin operators
in terms of Eq.~(\ref{eq:su2spexp}).
If a site index $i$ appears only once in $[\cdots]$,
every term in the expansion is equal to zero
because the average at site $i$ is evaluated as
$
\langle s_i^{x} \rangle_0=
\langle s_i^{y} \rangle_0=
\langle s_i^{z} \rangle_0=0.
$
Consequently, the cumulant is equal to zero.

Note that, in contrast to the Ising model,
a moment bond-diagram in which
an odd number of bonds meet at a site
can give nonzero value because
a product of three operators can give nonzero average, for example,
$\langle
s_i^{x} s_j^{x}\;
s_i^{y} s_j^{y}\;
s_i^{z} s_j^{z}
\rangle_0$.

Next, we prove Fig.~\ref{fig:bonddia1b}(e).
After Sec.~\ref{sec:prooflemma} is explained,
this property can be easily proved.
Let us think about
the mixed expansion pivoting on the bond in the center.
Then, $C_1$ is equal to zero because
this middle bond always unites two cycles.
Furthermore, $C_2$ in Eq.~(\ref{eq:c1plusc2}) is equal to zero
because every cumulant in it is equal to zero.

\section{simple FCM formulae in one dimension}
\label{app:freeene}

In one dimension, the equations in Sec.~\ref{sec:exFCM}
can be reduced to simpler forms as shown below.
From Eq.~(\ref{eq:excsum}), we can derive
\begin{eqnarray}
\langle {p}_{i,j}\rangle_\ell^\prime
& = &
\langle {p}_{i,j}\rangle_\ell- \langle {p}_{i,j}\rangle_{\ell-1}
-\langle {p}_{i-1,j-1}\rangle_{\ell-1} \nonumber\\
&&
\;\;+\langle {p}_{i-1,j-1}\rangle_{\ell-2}.
\label{eq:relation}
\end{eqnarray}
%
In other words, Eq.~(\ref{eq:excsum}) is inverted into Eq.~(\ref{eq:relation})
using {\it full perimeter lattice constants} explained
in Ref.~\onlinecite{DombDomb}.
Combining Eqs.~(\ref{eq:totcont}) and (\ref{eq:relation}),
contributions from small clusters are canceled out and
we obtain a simple formula,
\begin{equation}
\langle {p}_{i,i+x}\rangle'_{\le\ell}=
\sum_{j=1}^{\ell-x} \langle {p}_{j,j+x} \rangle_\ell
-\sum_{j=1}^{(\ell-1)-x} \langle {p}_{j,j+x} \rangle_{\ell-1}.
\end{equation}

There are formulae also for the free energy
and we have used it for calculating the specific heat of
the SU($\infty$) limit in Ref.~\onlinecite{FK1d}.
The free energy from the interaction part
for ${\cal H}_\ell^{(p)}$ is
$f_\ell:=-\beta^{-1} \ln\langle \exp(-\beta {\cal H}_\ell^{(p)})\rangle_0$.
The {\it net} contribution from the $\ell$-site cluster is defined as,
\begin{equation}
f'_{\ell}:= f_\ell-\sum_{\ell_1=1}^{\ell-1} (\ell_1+1) f'_{\ell - \ell_1}.
\label{eq:netf}
\end{equation}
The total net contribution
from clusters smaller than or equal to $\ell$ is defined as
\begin{equation}
f'_{\le\ell}:= \sum_{l=1}^\ell f'_{l}.
\label{eq:netftot}
\end{equation}
The simplified version of these relations also exists.
Equation (\ref{eq:netf}) is rewritten as\cite{DombDomb}
\begin{equation}
f'_{\ell}= f_\ell- 2 f_{\ell-1} + f_{\ell-2}.
\end{equation}
Finally,  Eq.~(\ref{eq:netftot}) is reduced to
\begin{equation}
f'_{\le\ell}= f_\ell-f_{\ell-1}.
\label{eq:freeenedrv}
\end{equation}
Here, $f'_{\le\ell}$ is correct up to $O[(\beta J)^{2\ell -1}]$.
Note that $f'_{\le\infty}= F_{\rm int}/N $,
where $F_{\rm int}:=-T \ln \langle \exp(-\beta {\cal H}^{(p)})\rangle_0$.
In fact, Eq.~(\ref{eq:freeenedrv}) is reduced to
$f'_{\le\infty} ={\rm d}F_{\rm int}/{\rm d}N$ in the $\ell\rightarrow\infty$
limit.

\section{ an example of the trace}
\label{sec:xtrace}

Let us consider a five-site system.
Then,
\begin{equation}
 {\rm Tr}^{(5)} \;1
= \sum_{\alpha_1=1}^n\cdots\sum_{\alpha_5=1}^n
\langle\alpha_1\alpha_2\alpha_3\alpha_4\alpha_5 |
\alpha_1\alpha_2\alpha_3\alpha_4\alpha_5 \rangle
=n^5.
\end{equation}
We show the calculation for $P=P_{1,4} P_{3,5} P_{1,2}$ below.
Here, $P_{i,j}$ exchanges $i$-th state and $j$-th state.
\begin{eqnarray}
&& {\rm Tr}^{(5)} P
\nonumber
\\
&=& \sum_{\alpha_1=1}^n\cdots\sum_{\alpha_5=1}^n
\langle\alpha_1\alpha_2\alpha_3\alpha_4\alpha_5 |
P|\alpha_1\alpha_2\alpha_3\alpha_4\alpha_5 \rangle
\nonumber
\\
&=& \sum_{\alpha_1=1}^n\cdots\sum_{\alpha_5=1}^n
\langle\alpha_1\alpha_2\alpha_3\alpha_4\alpha_5 |
\alpha_4\alpha_1\alpha_5\alpha_2\alpha_3 \rangle
\nonumber
\\
&=& \sum_{\alpha_1=1}^n\cdots\sum_{\alpha_5=1}^n
\delta_{\alpha_1,\alpha_4}
\delta_{\alpha_2,\alpha_1}
\delta_{\alpha_3,\alpha_5}
\delta_{\alpha_4,\alpha_2}
\delta_{\alpha_5,\alpha_3}
\end{eqnarray}
Let us consider a condition to give a nonzero contribution.
Here, $\delta_{\alpha_1,\alpha_4}$
requires $\alpha_1=\alpha_4$,
then $\delta_{\alpha_4,\alpha_2}$
requires $\alpha_4=\alpha_2$,
then $\delta_{\alpha_2,\alpha_1}$
requires $\alpha_2=\alpha_1$.
These are reduced to $\alpha_1=\alpha_4=\alpha_2$.
In fact, this procedure is equivalent to finding
a cyclic permutation $1\rightarrow4\rightarrow2\rightarrow1$.
By the same argument,
variables in another cyclic permutation $3\rightarrow5\rightarrow3$
also should be equal to each other, namely, $\alpha_3=\alpha_5$.
Hence,
there are only two independent variables in the multiple summation above.
Accordingly, we obtain
\begin{equation}
{\rm Tr}^{(5)} P =n^2.
\end{equation}
In other words,
permutation $(12345) \rightarrow (41523)$ is composed of
cycles $(142) \rightarrow (421),\; (35)\rightarrow(53)$,
and then, the number of independent variables is given by
the number of cycles, namely, two.

Let us operate one more exchange operator to $P$.
When $P_{1,2}$ is operated,
both 1 and 2 belong to a cycle $(142)$,
and this operation breaks this cycle.
That is, $P_{1,2} P$ has three cycles (1)(24)(35).
In the case of $P_{1,3} P$, however,
1 belongs to (142), and 3 belongs to (35),
and these cycles are united, namely,
$P_{1,3} P$ has only one cycle (15342).

\section{another definition of quasi-moments/cumulants}
\label{sec:otherdef}

In fact, a quasi-moment can be defined
without a prefactor in Eqs.~(\ref{eq:qmobtain})
and (\ref{eq:qmobtain2}).
In this case, Fig.~\ref{fig:bonddia3}(b) is equal to
a quasi-moment.
The generating function for this quasi-moments is
defined by a derivative as,
\begin{eqnarray}
\tilde{\cal G}_{\rm qm}&:=&
\frac{\partial}{\partial \lambda_{i,i+1}}
\frac{\partial}{\partial \lambda_{i+1,i+2}}
\cdots
\frac{\partial}{\partial \lambda_{j-1,j}}
\nonumber \\
&& \times \frac{\langle p_{i,j} \rangle}
{[P_{i,j} \, P_{i,i+1} P_{i+1,i+2} \cdots P_{j-1,j}]_{\rm s}}.
\end{eqnarray}
Consequently,
the generating function for the corresponding quasi-cumulants is
written as,
\begin{equation}
\tilde{\cal G}_{\rm qc}:= \ln \tilde{\cal G}_{\rm qm}.
\end{equation}
However, we have to integrate the quasi-moment generating function
$j-i-1$ times  to come back to the correlation
function. Hence, it is difficult to calculate general dependence
on $j-i$, and thus this definition is not suitable for our purpose.
This is the reason why we have adopted Eq.~(\ref{eq:defgqm}).

\section{amida-diagrams}
\label{sec:proofspht}

\begin{figure}[h]
\includegraphics[width= 7cm]{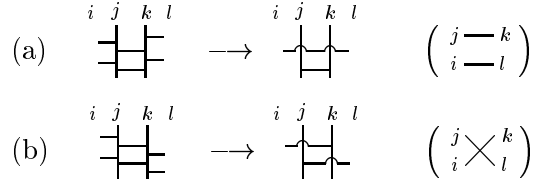}
\caption{A part of amida-diagrams.
The pairs of two bonds can be simplified.
Diagrams in parentheses represents
which site belongs to which cyclic permutation.
\label{fig:amida1st22}}
\end{figure}

First, let us think about moment amida-diagrams.
In order to give a nonzero contribution,
a configuration at a nearest-neighbor pair is topologically
equivalent to one of the two types shown in Fig.~\ref{fig:amida1st22}.
A pair of $P_{x,y}$ can be simplified
by exchanging subscripts $x$ and $y$ of operators
between them.
We can simplify left and right pairs of two bonds
to give the r.h.s$.$ of Fig.~\ref{fig:amida1st22}.
In the parentheses, we schematically represent
which sites are included in which cycle;
sites belonging to a cycle are connected by a line.

\begin{figure}[h]
\includegraphics[width= 7cm]{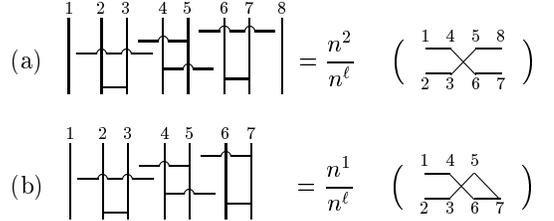}
\caption{ The simplified diagrams are made of these pieces.
For a guide to an eye, the pieces are separated by small spaces.
Diagrams in parentheses represent
which site belong to which cyclic permutation.
\label{fig:amida1st221}}
\end{figure}

In the moment versions of amida-diagrams in Fig.~\ref{fig:amida1st1},
we simplify bonds at every other intersite, namely,
$P_{1,2}$, $P_{3,4}$, \dots.
The simplified diagrams are made of the pieces in Fig.~\ref{fig:amida1st22}
and special pieces for the ends as shown in Fig.~\ref{fig:amida1st221}.
When the end pieces are absent,
any combination of the pieces in Fig.~\ref{fig:amida1st22}
gives two cycles.
When $\ell$ is an even number,
the end pieces are simple vertical lines, and do not do anything
to cycles.
Hence, the diagram has two cycles and gives $n^2/n^\ell$.
When $\ell$ is an odd number, one of the end pieces is not
a simple vertical line as shown in Fig.~\ref{fig:amida1st221}(b),
and it unites two cycles into one.
Therefore, the diagram has only one cycle and gives $n/n^\ell$.
The important point is that the amida-diagrams of odd $\ell$
are equal to those of $\ell+1$.

\begin{figure}[h]
\includegraphics[width= 6cm]{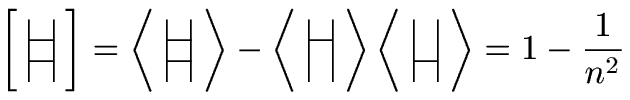}
\caption{Explicit calculation of the lowest order.
\label{fig:amida1st222}}
\end{figure}

Then, we can prove Fig.~\ref{fig:amida1st1}
using a mathematical induction.
In Fig.~\ref{fig:amida1st222},
we explicitly calculates the lowest order.
Next, let us consider the mixed expansion
in Fig.~\ref{fig:amida1st4} for the higher orders.
Suppose $\ell$ is an odd number.
The first and second terms cancel out each other
by the same argument
as noted in the context of
Eq.~(\ref{eq:lemma}) and Fig.~\ref{fig:upsilon0}.
If Fig.~\ref{fig:amida1st1} is true for the lower orders,
the sum of the third and fourth terms,
the sum of the fifth and sixth terms, \dots,
are equal to zero.
As a result, only the last term survives.
Since the moment amida-diagram in it is equal to unity,
we obtain $-(1-n^{-2})$.
Next, suppose $\ell$ is an even number.
The sum of the first and the second terms is
$\frac{n^2}{n^\ell} (1-\frac{1}{n^2})$.
The third term is
$-(1-\frac{1}{n^2})$ times $\frac{n}{n^{\ell-1}}$.
Hence, the first, second and third terms cancel out.
Furthermore,
if Fig.~\ref{fig:amida1st1} is true for the lower order,
the sum of the fourth and fifth terms,
the sum of the sixth and seventh terms, \dots,
are equal to zero.
Namely, again, only the last term survives.
Since the moment amida-diagram in it is equal to unity,
we obtain $(1-n^{-2})$.


\begin{thebibliography}{}

\bibitem{Domb3}
G.~S.~Rushbrooke, G.~A.~Baker,~Jr. and P.~J.~Wood,
in {\it Phase Transitions and Critical Phenomena},
edited by C.~Domb and M.~S.~Green
(Academic, London, 1974), vol.3, p.245.

\bibitem{oitmaa} J.~Oitmaa and E.~Bornilla,
Phys. Rev. B {\bf 53}, 14228 (1996).

\bibitem{Gelfand} M.~P.~Gelfand, R.~R.~P.~Singh, D.~A.~Huse,
J. Stat. Phys. {\bf 59}, 1093 (1990); M.~P.~Gelfand, R.~R.~P.~Singh,
Adv. Phys. {\bf 49}, 93 (2000).


\bibitem{FK1d}
 N.~Fukushima and Y.~Kuramoto,
J. Phys. Soc. Jpn. {\bf 71}, 1238 (2002).

\bibitem{Kubo}
R.~Kubo, J. Phys. Soc. Jpn.  {\bf 17}, 1100 (1962).

\bibitem{Singhcomment}
For this purpose,
an extrapolation of the FCM result and
a direct calculation of a few leading orders
have been done:
A.~B\"{u}hler, N.~Elstner and G.~S.~Uhrig,
Eur. Phys. J. B {\bf 16}, 475 (2000);
W.~O.~Putikka, M.~U.~Luchini and R.~R.~P.~Singh, Phys. Rev. Lett.
 {\bf 81}, 2966 (1998).

\bibitem{hand}D.~C.~Handscomb, Proc. Camb. Phyl. Soc.
 {\bf 60}, 115 (1964).
\bibitem{Chen}H.~H.~Chen and R.~K.~Joseph, J. Math. Phys.
 {\bf 13}, 725 (1972).

\bibitem{Fulde}
P.~Fulde,
in {\it Electron Correlations in Molecules and Solids},
(Springer-Verlag, 1995), p.82.

\bibitem{DombDomb}
C.~Domb,
in {\it Phase Transitions and Critical Phenomena},
edited by C.~Domb and M.~S.~Green
(Academic, London, 1974), vol.3, p.1.



\end{thebibliography}
\end{document}